\documentclass[aip,preprint]{revtex4-1}
\usepackage{hyperref}
\usepackage[english]{babel}
\usepackage{braket}
\usepackage{mathtools}
\usepackage{dsfont}

\begin{document}

\title{Pilot-wave quantum theory with a single Bohm's trajectory\\}
\date{\today}
\author{Francesco Avanzini}
\email[e-mail: ]{francesco.avanzini@studenti.unipd.it}
\author{Barbara Fresch}
\email[e-mail: ]{bfresch@ulg.ac.be}
\altaffiliation[Presently at the ]{D\'epartement de Chimie, Universit\'e de Li\`ege, Belgium}
\author{Giorgio  J. Moro}
\email[e-mail: ]{giorgio.moro@unipd.it}

\affiliation{Dipartimento di Scienze Chimiche, Universit\`a di Padova, via Marzolo 1, 35131 Padova, Italy}
\begin{abstract}
The representation of a quantum system as the spatial configuration of its constituents evolving in time as a trajectory under the action of the wave-function, is the main objective of the De Broglie-Bohm theory (or pilot wave theory). However,  its standard formulation is referred to the statistical ensemble of its possible trajectories. The statistical ensemble is introduced in order to establish the exact correspondence (the Born's rule) between the probability density on the spatial configurations and the quantum distribution, that is the squared modulus of the wave-function. In this work we explore the possibility of using the pilot wave theory at the level of a single Bohm's trajectory, that is a single realization of the time dependent configuration which should be representative of a single realization of the quantum system. The pilot wave theory allows a formally self-consistent representation of quantum systems as a single Bohm's trajectory, but in this case there is no room for the Born's rule at least in its standard form. We will show that a correspondence exists between the statistical distribution of configurations along the single Bohm's trajectory and the quantum distribution for a subsystem interacting with the environment in a multicomponent system. To this aim, we present the numerical results of the single Bohm's trajectory description of the model system of six confined rotors with random interactions. We find a rather close correspondence between the coordinate distribution of one rotor, the others representing the environment, along its trajectory and the time averaged marginal quantum distribution for the same rotor. This might be considered as the counterpart of the standard Born's rule when the pilot wave theory is applied at the level of single Bohm's trajectory. Furthermore a strongly fluctuating behavior with a fast loss of correlation is found for the evolution of each rotor coordinate. This suggests that a Markov process might well approximate the evolution of the Bohm's coordinate of a single rotor (the subsystem) and, under this condition, it is shown that the correspondence between coordinate distribution and quantum distribution of the rotor is exactly verified.
\end{abstract}

\maketitle

\section{Introduction}
Since its origin, quantum mechanics gave rise to vivid debates about the possible physical interpretations of its mathematical formalism. The widely accepted Copenhagen interpretation~\cite{neumann1955} associates the wave-function to the probability of the outcomes in the measure of observables of the quantum system by invoking a purely classical measurement apparatus. It represents an epistemological approach since it does not attribute physical meaning to the wave-function independently on the act of measure. In this framework the Born' rule identifies the probability density for a particular configuration of the constituents with the square modulus of the wave function.

In 1952 David Bohm elaborated a different interpretation~\cite{Bohm1, Bohm2} called  ``pilot wave theory'', previously suggested by De Broglie at the Solvay congress in the 1927~\cite{broglie1928}. Indeed it is called also ``De Broglie-Bohm theory''. The pilot wave theory assumes that the physical constituents of the quantum systems are particles possessing like in classical mechanics well defined coordinates, i.e., the configuration of the system. The wave function plays the role of a field, like the electromagnetic field but in the configuration space, that pilots the particle's motion. Thus, the particle's coordinates and the wave-function are supposed to be reality elements, in other words an ontological status is attributed to both of them. The particle's configuration together with the wave function represents the dynamical state of the system which evolves in time deterministically. In such a framework the system's physical properties are specified according to the particle's configuration without reference to the measurement. 

This description in terms of the time dependent wave-function and of the trajectory of particle's configuration together, is well defined formally but it leave open the issue whether a relation exists with other interpretations of quantum mechanics. In order to provide an answer, Bohm moved from a single trajectory picture of the reality to a representation with an ensemble of particle's configurations, i.e., a swarm of trajectories evolving in time under the action of the same pilot, that is the same wave-function. 
A density distribution is defined for such an ensemble to describe the probability density that the set of particles is in a given configuration. Bohm introduced the postulate that the distribution of the system's configuration at the initial time coincides with the square modulus of the wave-function~\cite{Bohm1}. Then one derives the equivalence at all times between the distribution of configurations  and the square modulus of the wave-function, this being equivalent to quantum hydrodynamics formulation of the Schroedinger equation by Madelung~\cite{madelung1927}. In this way, the agreement with the Born's rule of standard quantum mechanics is assured. 

On the other hand there are no evident reasons for supporting on a physical ground the equivalence between the square modulus of the wave-function and the density of system's configurations at the initial time. Bohm himself recognized the critical role of this assumption and argued that if this is not the case, then the randomness deriving from particle's interactions would enforce such a correspondence during the time evolution~\cite{bohm1953}. Such a point of view has been further developed recently by Valentini and coworkers with the objective of demonstrating that an initial arbitrary distribution on the configuration space relaxes in the time to the square modulus of the wave-function~\cite{valentini1991a, valentini1991b, towler2012}. A different procedure has been proposed by Durr and coworkers by introducing an effective wave function representative of a system interacting with the environment~\cite{durr1992}.

The De Broglie-Bohm theory found applications in different aspects of quantum physics: the most common examples concern the analysis of specific  phenomenas, as tunneling or scattering~\cite{norsen2013, norsen2014}, modeled by a wave packet motion of one particle systems.  These studies show that a simple picture of particle's trajectories can be derived for the quantum dynamics. In other research fields the pilot wave theory has mainly a computational role in order to reconstruct the time dependent wave-function from a collection of evolving system's configurations~\cite{lopreore1999, wyatt1999a, wyatt1999b}. Furthermore pilot wave theory has been considered as the appropriate framework to address the semi-classical approximation of quantum mechanics~\cite{garashchuk2014,garashchuk2002}. There are also attempts to use it as tool in the development of new multiscale procedures for large size systems, with traditional quantum computational approaches handling only a restricted member of degrees of freedom, while the remaining are treated according to classical formalisms~\cite{garashchuk2012}.

In this work we intend to explore the application of the pilot wave theory at the level of a single trajectory of the system's configuration. The underlying motivations derive from the recent investigations on single-molecule or single-spin observables~\cite{berezovsky2008, neumann2008}. This together with the efforts towards the realization of quantum computers based on nanostructures~\cite{biercuk2009, suter2008, nielsen2000}, calls for a representation of material systems according to a single realization of the quantum state. In this respect, the formal structure of the De Broglie-Bohm theory is well defined and self-consistent also when an unique wave-function and an unique time-dependent system's configuration are used to describe a particular realization of the quantum system. This point of view might be considered as the most natural way of interpreting the pilot wave theory without requiring any particular constraint on the distribution of initial configurations. Indeed, the original formulation of the Bohm theory includes implicitly the idea that a real system has always a well defined configuration, and we intend to explore the implications of this picture of the real systems through a single trajectory.

In this way, however, a major issue remains open: what is the connection with the Born's rule of standard Quantum Mechanics? More specifically, how to define from a single trajectory of the system's configuration a probability density on the particle's coordinates, which is a prerequisite before to establish a relation with the wave-function? One can exploit the analogy with classical statistical mechanics which introduces the equilibrium distribution by considering the density of phase space points along a single time dependent realization of the isolated system~\cite{khinchin}. Also in the case of the pilot wave theory with a single Bohm trajectory (i.e., the time  dependent system's configuration), one can define in an analogous way the equilibrium probability density on the set of particle's coordinates. But then, how to compare such an equilibrium distribution with the quantum distribution given as the square modulus of the wave function, which is intrinsically a time-dependent distribution? As a matter of fact the comparison becomes meaningful when the marginal distributions are considered for a subsystem interacting with a larger environment acting as thermal bath. Indeed, by employing the methods developed in a previous work~\cite{flureddensmatrix}, one can show that in such a case the fluctuations of the marginal quantum distribution become negligible. Our conjecture is that in this particular situation the marginal distributions obtained from the configuration distribution along a Bohm trajectory and from the wave-function tend to coincide. In order to provide evidences about this behavior, we shall present some computational results for a model system of several, randomly coupled, confined rotors. It should be evident that such a conjecture, if verified, plays the same role of the Born's rule in the Bohm analysis of the ensemble of trajectories.

Few attempts has been done to connect the Born's rule with a single Bohm's trajectory. Shtanov~\cite{shtanov1997} investigated the problem from the point of view of ergodicity. Very  recently, Philbin~\cite{philbin2014} considered a simple one dimensional system (an harmonic oscillator) in the presence of an external time dependent potential which mimics the position measurement. From a temporal sequence of these position measurements he obtains the same distribution given by the square modulus of the wave-function. In spite of the differences  on the employed model systems and on the type of dynamical regime, we share the same objective of developing the pilot wave theory for  a single realization of the quantum system. 

The paper is organized as follows. Since we consider a model system composed of several interacting components, statistical tools are required to analyze the quantum pure state represented by the wave-function. In the next section we introduce the Random Pure State Ensemble (RPSE) employed for the sampling of the wave-function, and we summarize its fundamental properties~\cite{RPSE, Emergence1, Beyond, flureddensmatrix}. Such a statistical ensemble allows one to evaluate the amplitude of fluctuations of the quantum observables (expectation values) with respect to their equilibrium values defined by time averages, and to estimate the behavior of fluctuation amplitudes in the thermodynamic limit for increasing size of the system. By recalling the results reported in Ref.~\cite{flureddensmatrix}, it is shown that the marginal distribution on a subsystem, as obtained from integration of the square modulus of the wave-function on the environment degrees of freedom, is characterized by fluctuations of vanishing amplitude for increasing size of the environment. Therefore the subsystem is described by a nearly stationary marginal quantum distribution if the environment is large enough, and it can be well approximated by its time average. In the following Section~\ref{1bohm} the standard form of the pilot wave theory is summarized, and the procedure for generating the single Bohm's trajectory is illustrated. In Section~\ref{numF} the model system of six interacting confined rotors is used to verify our conjecture. First the model system is described in detail together with the numerical procedures employed for the calculation of time dependent properties. Then the main results are illustrated in relation to: 1) the nearly stationarity of the marginal quantum distribution of one rotor, the other five rotors constituting the environment, 2) the randomly fluctuating behavior displayed by the evolution of the Bohm's coordinate of one rotor with the corresponding loss of correlation with the time, 3) the close correspondence between the marginal quantum distribution and the distribution of Bohm's coordinate of the subsystem, which provides the computational evidence of our conjecture. In Section~\ref{just} we show that the conjecture is exactly verified if the Bohm coordinate of the subsystem behaves like an independent Markov stochastic variable, as partially suggested by the numerical results. In the final Section~\ref{concl} the general conclusions are drawn by focusing on the implications of our work.


\section{Statistics of quantum pure states\label{stat}}
In this section we present the statistical description of quantum pure states to be employed in the analysis of single Bohm trajectories. The need of a statistics of pure states, that is of quantum states described by a wave function belonging to  the  proper Hilbert space for an isolated (closed) system, arose mainly from the efforts of demonstrating the typicality of quantum observables~\cite{popescu2006, goldstein2006}. On the other hand, well defined statistical rules are required for sampling the initial wave function whenever the quantum dynamics is examined without particular a priori choices of the initial state.  We stress  that in quantum mechanics the condition of isolated system is more stringent than in classical mechanics: entanglement would keep the system connected to his environment even though there is no energy exchange between them.

In standard quantum mechanics~\cite{cohen} the wave function $\ket{\Psi(t)}$ is an Hilbet $\mathcal{H}$ space's element  representing  the state of the isolated system, which evolves in time through the Schr\"odinger equation from a given initial state $\ket{\Psi(0)}$,
\begin{equation}
\ket{\Psi(t)}=e^{-\imath \hat{H}t / \hbar}\ket{\Psi(0)}
\label{soluzioneEqS}
\end{equation}
$\hat{H}$ being the Hamiltonian of the system. The time evolution $A(t)$  of a  generic physical property   described by self-adjoint operator $\hat A$  is determined by the expectation value,
\begin{equation}
A(t)=\braket{\Psi(t)|\hat A|\Psi(t)}=\text{Tr}\Big\{\hat{A}\hat{\varrho}(t)\Big\},\label{aspval}
\end{equation}
where $\hat{\varrho}(t)$ is the  density matrix operator for the pure state
\begin{equation}
\hat{\varrho}(t)=\ket{\Psi(t)}\bra{\Psi(t)}\label{dm}.
\end{equation}
An expectation value $A(t)$  is usually interpreted as the mean value of a infinite number of measures of the observable at time $t$.
Like in classical statistical mechanics~\cite{khinchin}, we identify the equilibrium value of observable $A$ with the asymptotic time average of the expectation value:
\begin{equation}
\overline{A(t)}:=\lim_{T\rightarrow +\infty}\frac{1}{T}\int_{0}^{T}\mathrm{d}t\text{ }A(t)=\text{Tr}\Big\{\hat{A}\overline{\hat\varrho(t)}\Big\}\label{eqval},
\end{equation}
where $\overline{\hat\varrho(t)}$ is the time average of the density matrix.

If we are interested to properties of a subsystem, we can imagine to partition the isolated system: the subsystem $S$, also denoted as system in the following, and the environment $E$ for the remaining part of the isolated system. Correspondingly the overall Hilbert  space $\mathcal{H}$  is factorized into the Hilbert spaces $\mathcal{H}_{S}$  of the system and  $\mathcal{H}_{E}$ of the environment, $\mathcal{H}=\mathcal{H}_{S}\otimes\mathcal{H}_{E}$. In this situation the observable $a(t)$ of interest, i.e., a property of the subsystem, is represented by the expectation value of an operator $\hat {a}\otimes\hat{\mathds{1}}_E$. 
The reduced density matrix operator~$\hat{\sigma}(t)$, as obtained by partial trace $\text{Tr}_{E}$  over the environment states of the pure state density operator,
\begin{equation}
\hat{\sigma}(t):=\text{Tr}_{E}\big\{\hat{\varrho}(t)\big\},
\end{equation}
allows the calculation of this expectation value within the subsystem Hilbert space
\begin{equation}
a(t)=\text{Tr}\Big\{\big(\hat {a}\otimes\hat{\mathds{1}}_E\big)\hat{\varrho}(t)\Big\}=\text{Tr}_S\Big\{\hat {a}\ \hat{\sigma}(t)\Big\}.\label{expvalsub}
\end{equation}
Its equilibrium value $\overline{a(t)}$ is defined again by time averaging and it can be evaluated like in  Eq.~\eqref{eqval} by means of the time average  $\overline{\hat{\sigma}(t)}$ of the reduced density matrix.

In order to formulate a statistical description of quantum pure states, a finite set of parameters identifying the instantaneous wave function has to be selected, very much like for the phase space of classical statistical mechanics. This requires the confinement of the wave function to a finite dimensional subspace of $\mathcal{H}$, say a $N$-dimensional subspace $\mathcal{H}_N$ in the following called as active space. To select the active space, it is convenient to resort to the orthonormal eigenstates $\ket{E_{k}}$ of the Hamiltonian:
\begin{equation}
\hat{H}\ket{E_{k}}=E_{k}\ket{E_{k}}.\label{tiSE}
\end{equation}
Like in previous works~\cite{RPSE, Emergence1, Emergence2, Beyond}, we shall employ the following  type of active space
\begin{equation}
\mathcal{H}_{N}:=\Biggl\lbrace\bigoplus_{k=1}^N\ket{E_k} \text{ with } E_N<E_{max}<E_{N+1}\Biggr\rbrace,
\label{activespace}
\end{equation}
that is the subspace due to eigenstates with eigenvalues smaller than $E_{max}$. The energy cutoff $E_{max}$ is the only parameter required for the identification of this active space, and it has been shown that  in the limit of macroscopic systems $E_{max}$ represents the internal energy~\cite{Beyond}.

It should be mentioned that one can employ an alternative active space  by introducing also a low energy cutoff $E_{min}$, like in the definition of microcanonical density matrix of standard quantum statistical mechanics~\cite{huang1987}. In this way, however, one has to manage two different cutoff parameters and, furthermore, no direct relation exists between the lower energy cutoff $E_{min}$ and thermodynamic properties~\cite{Beyond}. 

The wave function $\ket{\Psi(t)}$ at a given time is then conveniently specified as a linear combination of the basis elements $\ket{E_{k}}$ of the active space through its expansion coefficients $c_k(t)$ or, equivalently, through the sets of populations $(P_1, P_2, \dots, P_{N})$ and of phases $\alpha(t)=(\alpha_1(t),\alpha_2(t), \dots,\alpha_N(t))$ obtained from the polar representation of the expansion coefficients
\begin{equation}
c_k(t):=\braket{E_k|\Psi(t)}=\sqrt{P_{k}}e^{-\imath\alpha_{k}(t)}.
\end{equation}
with a linear time dependence of the phases: $\alpha _k(t)=\alpha _k(0)+E_kt/\hbar$. Because of the normalization condition,
\begin{equation}
\braket{\Psi(t)|\Psi(t)}=\sum_{k=1}^{N}P_k=1,
\end{equation}
only $(N-1)$ populations are independent, say the set  $P=(P_1, P_2, \dots, P_{N-1})$. Therefore a bijection exists between  the normalized wave function and the ensemble of populations $P$ and of phases  $\alpha(t)$, with each particular set of these $(2N-1)$ real parameters corresponding to a specific wave function $\ket{\Psi(t)}$. 
In other words, all the pure states of the active space can be imagined like unit vectors drawing  an unit sphere in a $2N$-dimensional Euclidean space~\cite{RPSE}.

Because of the choice of expanding the wave function along the Hamiltonian eigenstates, the phases are the only dynamic variables of the system, while the populations represent  the constants of motion. Correspondingly it is easily shown that equilibrium properties $\overline{A(t)}$  of Eq.~\eqref{eqval} depend on populations only. Indeed, under the condition of rational independence of Hamiltonian  eigenvalues, meaning that equation $\sum_{k=1}^{N}n_{k} E_{k}=0$ for integer $n_k$ has only the trivial solution $n_k=0$ $\forall k$, the equilibrium density matrix is diagonal
 with the populations as diagonal elements~\cite{RPSE},
\begin{equation}
\overline{\hat{\varrho}(t)}=\sum_{k=1}^{N}\ket{E_{k}}P_{k}\bra{E_{k}}\equiv\overline{\hat{\varrho}}^{P},\label{eqdm}
\end{equation}
where we have introduced the symbol $\overline{\hat{\varrho}}^{P}$ to highlight the dependence of equilibrium density matrix on populations only. We emphasize that the condition of rational independence is not too restrictive because of the contribution of random interactions, typical of material systems, 
leading to energy eigenvalue distribution having, at least partially, a random  character~\cite{randommatrix}. Therefore, according to  Eq.~\eqref{eqval}, also the equilibrium value of a generic observable depends on populations only,
\begin{equation}
\overline{A(t)}=\text{Tr}\Big\{\hat{A}\overline{\hat{\varrho}}^{P}\Big\}\equiv\overline{A}^{P}, 
\end{equation}
as stressed by the symbol $\overline{A}^{P}$.

In this way, the parametric dependence of the equilibrium properties on the populations results to be evident. On the other hand, there are no empirical methods leading to a complete characterization of the initial state   $\ket{\Psi(0)}$ and, therefore, also of the populations. This implies that populations can be characterized only on a statistical ground by selecting the ensemble for their probability distribution. The absence of privileged directions for $\ket{\Psi(0)}$ within the unit sphere, leads quite naturally to a purely random choice for the ensemble of pure states. In Ref.~\cite{RPSE} the Random Pure State Ensemble (RPSE) for populations has been characterized from the geometrical analysis of the measure on the unit sphere, so deriving the probability density on the $(N-1)$ independent populations $P$
\begin{equation}
p_{\text{RPSE}}(P)=(N-1)!.\label{RPSEd}
\end{equation}
Such a probability density allows the explicit calculation of the ensemble  average of equilibrium properties,
\begin{equation}
\Big\langle \overline{A}^{P}\Big\rangle:=\int \mathrm{d}P_{1}\dots\mathrm{d}P_{N-1} \text{ }\text{ } \overline{A}^{P}\text{ } p_{\text{RPSE}}(P_{1},\dots, P_{N-1}),
\end{equation}
which can be interpreted as the average of $\overline{A}^{P}$ amongst random realizations of the initial pure state $\ket{\Psi(0)}$. Notice that integration domain on populations is bounded by constraints $0\le P_k \le 1 , \text{ } \forall k=1,2, \dots N$. In the following we shall employ the bracket  $\Big\langle   \dots \Big\rangle$ to denote the RPSE average of a function of populations. 

In order to recover also the macroscopic description of the system, one should consider the  equilibrium energy, $\overline{H}^P=\sum_{k=1}^{N}P_kE_k$, and Shannon's entropy~\cite{shannon} with respect to the populations, $S^P=-k_{B}\sum_{k=1}^{N}P_k\ln P_k$. Their RPSE average are associated respectively to the thermodynamical internal energy, $U:=\langle\overline H^P\rangle$, and to the thermodynamical entropy, $S:=\langle S^P\rangle$, both being functions of $E_{max}$. By eliminating the $E_{max}$ dependence between functions 
$U(E_{max})$ and $S(E_{max})$, one recovers  the thermodynamical state function $S(U)$  and the temperature as well from its derivative $1/T=dS/dU$. In this framework, by considering  the system as an ensemble of $n$  distinct components, like molecules in material systems, one can define the thermodynamic limit for $n\to \infty$  at a given temperature~\cite{Beyond}. The thermodynamic limit requires the tensorial product of the Hilbert spaces of all the distinct components, and this implies an exponential growth  of the dimension $N$ of the active space $\mathcal{H}_N$ with the number $n$ of components~\cite{flureddensmatrix}. Finally, in the same limit, the RPSE average of the equilibrium reduced density matrix $\Big\langle\overline{\hat{\sigma}}^{P}\Big\rangle$ of a subsystem having weak enough interactions with the environment, takes the canonical form 
\begin{equation}
\Big\langle\overline{\hat{\sigma}}^{P}\Big\rangle=\frac{e^{-\hat{H}_{S}/k_{B}T}}{\text{Tr}_{S}\Big\{e^{-\hat{H}_{S}/k_{B}T}\Big\}}\label{averagerdm},
\end{equation}
where $\hat{H}_{S}$ is the Hamiltonian of the subsystem~\cite{Beyond}. 

The RPSE statistics allows the quantitative analysis of typicality~\cite{Beyond} of an equilibrium property $\overline{A}^{P}$ by evaluating the thermodynamic limit of its square variance within the ensemble,
\begin{equation}
\lim_{n\rightarrow\infty}\bigg\langle\Big(\overline{A}^P-\Big\langle\overline{A}^{P}\Big\rangle\Big)^{2}\bigg\rangle.\label{typicality}
\end{equation}
Typicality of property $\overline{A}^{P}$ is assured if this limit vanishes, this implying that the value of $\overline{A}^{P}$ in a realization of the pure state is independent of the set of populations, as long as its deviation from the ensemble average $\Big\langle \overline{A}^{P}\Big\rangle$ tends to vanish. In other words, property $\overline{A}^{P}$ is typical in the meaning that it is nearly independent of the particular realization of the pure state.

Furthermore, RPSE ensemble allows the quantitative analysis not only of typicality of an observable, but also of its time fluctuations which are of primary importance for the objectives of this work. In order to characterize the amplitude of fluctuations of $A(t)$ during its time evolution, we consider the equilibrium value, i.e., the time average, of the square of deviations $\Delta A(t):=A(t)-\overline{A}^{P}$ from the time average
\begin{equation}
\overline{(\Delta A)^{2}}^{P}:=\overline{\Big(A(t)-\overline{A}^P\Big)^2}
\end{equation}
that, like all the equilibrium properties, depends on the population set. The population average within RPSE provides an estimate $\Big\langle\overline{(\Delta A)^{2}}^{P}\Big\rangle$ of squared fluctuations which is independent of the particular realization of the pure state~\cite{bartsch2009, flureddensmatrix} and reads
\begin{equation}
\Big\langle\overline{(\Delta A)^{2}}^{P}\Big\rangle+
\bigg\langle\bigg(\overline{A}^P-\Big\langle\overline{A}^{P}\Big\rangle\bigg)^{2}\bigg\rangle
=\frac{D_{2}(\hat A)}{N+1}
\label{T1},
\end{equation}
where the second term at the left hand side describes the typicality of equilibrium property $\overline{A}^{P}$ as previously discussed. At the right hand side, $N$ is the dimension of the active space $\mathcal{H}_N$, while $D_{2}(\hat A)$ represents the squared spectral variance of the operator $\hat A$, $D_{2}(\hat A)=\sum_{k=1}^N (\lambda_k-D_1(\hat A))^2$, where $\{\lambda_k\}$ is the ensemble of eigenvalues of $\hat A$ in $\mathcal{H}_N$ and $D_1(\hat A)=\sum_{k=1}^N \lambda_k/N$ is the eigenvalue average. Such a relation connects the statistical properties of the expectation value $A(t)$, at the left hand side of the equation, to the spectral properties of the operator $\hat A$, on the right hand side of equation. If operator $\hat A$ has a bounded spectrum, then $D_{2}(\hat A)$ is finite and  in the thermodynamic limit,  $n\rightarrow +\infty$,  the right hand side of Eq.~\eqref{T1} vanishes because of the exponential growth with $n$ of the active space dimension $N$. Correspondingly also both terms at the left hand side of Eq.~\eqref{T1} vanish since they are non negative
 \begin{equation}
\lim_{n\rightarrow +\infty}\bigg\langle\bigg(\overline{A}^P-\Big\langle\overline{A}^{P}\Big\rangle\bigg)^{2}\bigg\rangle=\lim_{n\rightarrow +\infty}\Big\langle\overline{(\Delta A)^{2}}^{P}\Big\rangle=0.
\end{equation}
Thus, in the thermodynamic limit, both typicality and the vanishing of fluctuations  are assured for bounded operators.  Outside the thermodynamic limit, for finite but large enough isolated quantum systems a nearly stationarity $A(t) \simeq \overline{A}^{P}$ is predicted. Furthermore, we note that in this conditions the expectation value $A(t)$ is nearly equal to the thermodynamic value $\Big\langle\overline{A}^{P}\Big\rangle$ because of typicality: $A(t)\simeq\Big\langle\overline{A}^{P}\Big\rangle$.

These results for typicality and fluctuation amplitude of bounded operators can be applied to the reduced density matrix of a subsystem of an isolated system. In particular, as shown in detail in Ref.~\cite{flureddensmatrix},  the following condition for the expectation value $a (t)$ of subsystem operator $\hat a$ derives from Eq.~\eqref{T1}, 
\begin{equation}
\begin{split}
\Big\langle\Big(\overline{a}^P-\big\langle&\overline{a}^{P}\big\rangle\Big)^{2}\Big\rangle+\Big\langle\overline{(\Delta a)^{2}}^{P}\Big\rangle\\
&\leq\frac{\text{Tr}_{S}\Big\{\hat{a}^{2}\Big\langle\overline{\hat{\sigma}}^{P}
\Big\rangle\Big\}-\text{Tr}_{S}\Big\{\hat{a}\Big\langle\overline{\hat{\sigma}}^{P}\Big\rangle\Big\}^{2}}{N+1}.
\end{split}
\label{T2}
\end{equation}
In the thermodynamic limit the ensemble average of the reduced density matrix tends to the canonical form Eq.~\eqref{averagerdm} and, therefore, the right hand side vanishes because of the active space dimension $N$ of  at the denominator. Then both typicality and the vanishing of fluctuations are recovered like in Eq.~\eqref{T1} for bounded operators, but now for a generic operator $\hat a$ of the subsystem. For finite but large enough isolated systems this implies that subsystem observables are nearly stationary,
\begin{equation}
a(t) \simeq \overline{a}^{P}
\label{circa}
\end{equation}
that is, their time dependent deviations from the equilibrium values is negligible.

As an application of the previous analysis, we examine the statistical distribution on the coordinates $q_S$ for the subsystem degrees of freedom. In standard quantum mechanics the wave function allows the calculation of the time dependent distribution on the ensemble of coordinates $q=\{q_k\}_{k=1,\dots,n}$ of the isolated system with $n$ degrees of freedom through the probability density
\begin{equation}
p(q,t)=\big|\Psi(q, t)\big|^2
\end{equation}
with a parametric dependence on the initial pure state determining the time dependent wave function. Once the subsystem $S$, and the environment $E$ as well, has been selected, the isolated system coordinates can be identified with the ensemble $q=(q_S,q_E)$ of subsystem coordinates $q_S$  and of coordinates $q_E$ for  the environment degrees of freedom. Then, by integration on the environment coordinates, the marginal distribution on the subsystem degrees of freedom is recovered
\begin{equation}
p^S(q_S,t):=\int \mathrm{d}q_{E}\text{ }p(q_S,q_E,t).
\label{pista}
\end{equation}
Like for any time dependent observable, the time average defines the corresponding equilibrium property, in this case the equilibrium distribution
\begin{equation}
p^{ S, eq}(q_{S}):=\overline{p^S(q_{S},t)},
\label{pmedia}
\end{equation}
where the reference to the parametric dependence on population set  $P$  has been omitted for the sake of a compact notation.

Let us consider now an orthonormal basis $\{\ket{\varphi_{m}}\}$ for the subsystem Hilbert space $\mathcal{H}_{S}$, and its representation $\{\varphi_m (q_S)\}$ as explicit functions of subsystem coordinates $q_S$. For any  set of $q_S$ values, we can define the following operator 
\begin{equation}
\hat{a}(q_{S}):=\sum_{m,m{'}}\ket{\varphi_{m}}\varphi_{m}^{*}(q_{S})\varphi_{m{'}}(q_{S})\bra{\varphi_{m{'}}}\label{opdp},
\end{equation}
where  its operator nature is determined by the 
bras $\bra{\varphi_{m{'}}}$ and kets  $\ket{\varphi_m}$ on the r.h.s.. One can easily verify that its $q_S$-dependent expectation value supplies the subsystem marginal probability density calculated at $q_S$
\begin{equation}
p^S(q_{S},t)=\text{Tr}_S \Big\{\hat {a}(q_S)\hat{\sigma}(t)\Big\}.\label{expdp}
\end{equation}
In this way, the marginal distribution can be interpreted as expectation value of a subsystem operator, which is characterized by typicality and absence of fluctuations in the thermodynamic limit in agreement with the previous conclusions. Outside the thermodynamic limit, but for large enough isolated systems, negligible contributions of fluctuations about the time average Eq.~\eqref{pmedia} are expected like in Eq.~\eqref{circa}, 
\begin{equation}
p^S(q_{S},t) \simeq p^{ S, eq}(q_{S})\label{appPDsub},
\end{equation}
so that the subsystem is characterized by a nearly time independent marginal distribution.

In conclusion, as long as expectation values or, equivalently, marginal distributions derived from the wave function are employed to describe a subsystem which is part of a much larger isolated system, the time evolution of the subsystem appears to be secondary.
As a matter of fact the environment quenches the dynamics of these subsystem properties.
In a classical world this would correspond to a picture of motionless subsystems, like molecules in material systems, without fluctuations in the thermodynamic limit. Such a stationarity derives from the fact that the expectation value is not a directly observed physical  property, but an average of infinite measures of the physical property. Nonetheless, the expectation value is the standard tool supplied by quantum mechanics for the description of the time evolution of physical properties, tool which displays stationarity  in the thermodynamic limit. It should then be  useful to explore the Bohm theory by  looking for alternative tools able to capture the fluctuation dynamics of parts, like molecules, of a larger isolated system.


\section{Pilot-wave theory\label{1bohm}}
Pilot-wave theory is a formulation of quantum mechanics firstly proposed by de Broglie~\cite{broglie1928,pilotwave} and afterwards rediscovered and fully developed by David Bohm in 1952~\cite{Bohm1, Bohm2}. It has several advantages with respect to more traditional approaches to quantum mechanics, since it leads to a full characterization of system constituents as particles having well defined  geometrical coordinates $Q(t)$ that evolves in time in a deterministic way. Furthermore, it allows a description of measurement processes  without the need of the  wave function's collapse. Beyond these evident benefits, we think that pilot-wave theory is important also because it allows a representation of subsystem dynamics which overcomes the stationarity found for expectation values  in the thermodynamic limit .

Given the polar representation of the wave function, $\Psi(q, t)=R(q, t) e^{\imath {S(q, t)}/{\hbar}}$, from the Schr\"odinger equation one can derive the time evolution equations for the amplitude $R(q,t)$ and for the quantum phase $S(q,t)$.  The equivalence between the latter equation and the classical  Hamilton-Jacobi equation is the 
starting point of Bohm's analysis, which leads to a rate equation for particle positions $Q(t)$ controlled by the phase $S(q,t)$ to be interpreted as a time dependent field. On this basis, Bohm formulated his theory that can be summarized by the following three main assumptions.

The first concerns the wave function $\Psi (q,t)$ representing a field which pilots the system's constituents (particles), and evolving in time according to standard   Schr\"odinger equation independently of particle positions 
\begin{equation}
\imath  \hbar \frac{\partial\Psi (q,t) }{\partial t}=\hat H \Psi(q, t).\label{shcrodingereq}
\end{equation}

Secondly, the instantaneous configuration of the system is specified through the set $Q(t)=\{Q_{k}(t)\}_{k=1,\dots,n}$ of coordinates of all its particles. Like in classical mechanics they evolve in time along a trajectory satisfying the following rate equation in the case of particle with mass $m_k$ 
\begin{equation}
m_k\frac{\mathrm{d}Q_k(t)}{\mathrm{d}t}=\frac{\partial S(q,t)}{\partial q_k}\bigg|_{q=Q(t)}\label{bohmeq}.
\end{equation}
Thus, once the wave function $\Psi (q,t)$ is provided at all times and if the initial configuration $Q(0)$ is known, in principle one can derive the trajectory $Q(t)$ for the evolution of the system configuration. A physical observable is represented simply by a function $A(Q)$ of the system configuration and its value $A(Q(t))$ at time $t$ is evaluated from the corresponding coordinates $Q(t)$. The quantum nature of system dynamics arises from the pilot role of the wave function on the coordinate evolution, generating trajectories in general different from those predicted by classical mechanics. 

The third assumption concerns the correspondence with the predictions of ordinary quantum mechanics as specified by the probability density $p(q,t)=\big|\Psi(q, t)\big|^2$ on the coordinates. This calls for a statistical description of the system
configurations by imagining a swarm  of trajectories generated by different initial conditions $Q(0)$, but with the same wave function as pilot agent.  Let us introduce the probability density $\rho _0 (q)$ for the initial coordinates $Q(0)$. Then we can define the probability density $\rho(q,t)$ of system coordinates at a generic time $t$,
\begin{equation}
\rho (q,t):=\int  \mathrm{d}q_0 \rho _0(q_0) \delta (q-Q(q_0;t)) ,\label{distswzz}
\end{equation} 
on the basis of trajectories $Q(q_0;t)$ starting at  $Q(q_0;0)=q_0$, always with the same pilot wave function. In this way, a meaningful comparison can be made between $\rho (q,t)$ and $p(q,t)$ since both are probability densities on the same variables. The identity of the initial coordinate distributions and of the initial quantum probability distribution, 
\begin{equation}
\rho(q,0)=\rho _0 (q)=p(q,0),
\label{Born0}
\end{equation}
is the third assumption of the Bohm's theory. This condition, known also quantum equilibrium~\cite{durr1992}, assures the equivalence between coordinates distribution and the squared norm of the wave function at all times~\cite{Bohm1}
\begin{equation}
\rho(q,t)=p(q,t)=\big|\Psi(q, t)\big|^2,
\label{Bornt}
\end{equation}
that is, the Born's rule for the correspondence between wave function and coordinate probability distribution. In this way, the same predictions are recovered from Bohm's theory and from traditional quantum mechanics. For instance the average of $A\big(Q(t)\big)$ on the ensemble  of trajectories, 
\begin{equation}
\int  \mathrm{d}q \text{ } A(q) \rho (q,t)  =A(t),
\end{equation}
becomes equivalent to the expectation value $A(t)$ of Eq.~\eqref{aspval}  for the operator $\hat A$ given as function $A(q)$.

It should be mentioned that distribution Eq.~\eqref{distswzz} on the ensemble of Bohm's trajectories can be specialized to the subsystem by integration on environment coordinates,
\begin{equation}
\rho^S (q_S,t):=\int  \mathrm{d}q_E\text{ }\rho (q_S,q_E,t) .
\end{equation} 
Such a reduced distribution is equivalent to the quantum marginal probability density Eq.~\eqref{pmedia}  provided that the quantum equilibrium condition Eq.~\eqref{Bornt} is satisfied,
\begin{equation}
\rho^S (q_S,t)=p^S(q_S,t) \simeq p^{S,eq}(q_S),
\end{equation} 
with stationarity holding in the absence of fluctuations for large enough systems.

Despite the original aim of Bohm's theory to overcome the methodological flaws of a traditional quantum formulation, in order to ensure the agreement with  predictions of quantum theory, it introduces a new controversial issue in relation to the third assumption and the role of Born's rule. First of all, there are not clear and evident justifications of the initial equivalence Eq.~\eqref{Born0} between coordinate distribution and squared modulus of the wave function. Such a issue has been tackled several times~\cite{bohm1953, durr1992, towler2012} with different proposals for the mechanisms ensuring the relaxation towards quantum equilibrium Eq.~\eqref{Bornt} if initially Born's rule is not satisfied. However, we think that the same idea of a swarm of trajectories rises some methodological criticisms. Indeed, by adopting a realistic point of view, a particular realization of the quantum system should be described by a single trajectory, while the swarm of trajectories should represent an ensemble of realizations of the system in correspondence of different initial configurations $Q(0)$.

We intend in this work to explore the Bohm's theory at the level of the single trajectory representation of the quantum system. Such an approach is equivalent to the Bohm's theory as long as the correspondence with standard quantum theory through Born's rule is not considered.  In this way, two reality elements describe completely the system's state: the coordinates and the wave function field, $\big(Q(t), \Psi(q,t)\big)$. 
The state evolution is strictly deterministic according to equation:
\begin{equation} 
\left(
\begin{aligned}
&\frac{\mathrm{d}Q(t)}{\mathrm{d}t}\\
&\frac{\partial\Psi (q,t)}{\partial t}
\end{aligned}
\right)=X\big(Q(t),\Psi(q,t) \big),
\label{Cauchy}
\end{equation} 
with a time independent vector field $X\big(Q, \Psi \big)$ derived from Schr\"odinger equation Eq.~\eqref{shcrodingereq} and from Bohm equation Eq.~\eqref{bohmeq}.
From a mathematical point of view, the system state is represented by an element of the union of the configuration space and of the  Hilbert space, with a time evolution described by  the Cauchy problem Eq.~\eqref{Cauchy}  having an unique solution for given initial configuration $Q_0\equiv Q(0)$ and initial wave function $\Psi _0(q) \equiv \Psi (q,0)$.

In order to recover a probabilistic description from a single Bohm's trajectory, one has necessarily to resort to the statistical sampling of the coordinates during their time evolution, like in ergodic theory of classical statistical mechanics~\cite{khinchin}. As long as such a sampling represents overall effects of system evolution, it is an equilibrium property which should in general depend on the constants of motion, that is the populations $P$ determining the pilot wave function.  The probability density on coordinates $q$ extracted from the sampling of a single trajectory $Q(t)$ will be denoted as $w^{eq} (q)$, keeping implicit the reference to the parametric dependence on the population set to deal with a more compact notation. In order to perform a meaningful comparison with the distribution obtained from the wave function, we shall consider in a multicomponent system the probability density $w^{S,eq}(q_S)$ on the subsystem $S$ described by coordinates $Q_S$.  As discussed in the previous section in relation to Eq.~\eqref{appPDsub}, if the isolated system is large enough, the quantum probability density of the subsystem  $p^S(q_{S},t)$ have negligible fluctuations and it can replaced by its equilibrium form $p^{ S, eq}(q_{S})$, that is a time independent distribution.  In other words, by examining a part of a much larger system, the quantum distribution can be described by  $p^{ S, eq}(q_{S})$, that is a time independent function like the coordinate distribution $w^{S,eq}(q_S) $ obtained from a single Bohm's trajectory and, therefore, a meaningful comparison between them can be done. This is the objective of the calculations in a model system reported in the next section.


\section{Bohm's trajectories in a multi-particle model system \label{numF}}
In order to compare the single Bohm's trajectory and the quantum distribution function, we have examined the dynamical behavior of a model system of six confined rotors  interacting through random potentials. The numerical calculations done for a typical situation clearly show that a correspondence exists between the Bohm's coordinate distribution $w^{S,eq} (q_S) $ and the equilibrium quantum distribution $p^{ S, eq}(q_{S})$ for a rotor subsystem. In this section, after the presentation of the model system, we discuss the numerical methods employed for the calculation of the relevant properties and we illustrate the most relevant results.

\subsection{The model system\label{system}}
We shall consider a system of $n=6$ identical but distinguishable particles with mass $m$ that move on a ring of constant radius $R$. Such a system is equivalent to $n$ planar rotors described by the set of angles $q=\{q_1,q_2,\dots, q_n\}$, each of them having an inertia momentum $I=mR^2$. A physical realization of the system could be an ensemble of methyl groups rotating about their $\text{C}-\text{CH}_3$ bonds. The Hilbert space $\mathcal{H}_i$ for the $i$-th rotor is the ensemble of periodic functions of the angular coordinate $q_i$, whose Fourier representations can be generated by means of the following orthonormal basis set
\begin{equation}
\chi_{j}(q_i)=\frac{e^{\imath j q_i}}{\sqrt{2\pi}}, 
\label{basis}
\end{equation}
with integer values for $j$ index. The tensor product of the Hilbert spaces of each rotor, $\mathcal{H}=\mathcal{H}_1\otimes\mathcal{H}_2\otimes\text{\dots} \otimes\mathcal{H}_n$,  identifies the Hilbert space for the overall system. Such a model system will be described by means of the following Hamiltonian
\begin{equation}
\hat H=\hat H^{(0)} + V^{(r)}=\sum_{i=1}^n \hat H_i^{(0)} +  {V}^{(r)},
\label{hamiltonian}
\end{equation}
where $\hat H_i^{(0)}$ is the single particle Hamiltonian, while $ V^{(r)}$ is an interaction potential of random type.
For the single particle Hamiltonian we use the model of a planar rotor confined by a cosine potential with minimum at $q_i=\pi$:
\begin{equation}
\hat H_i^{(0)}=-\frac{\hbar^2}{2I }\frac{\partial^2}{\partial q^2_i} 
+\frac{u}{2 } (1+\cos q_i) ,
\label{Hrotor}
\end{equation}
the parameter $u$ representing the energy barrier at $q_i=0$. In the following the parameter $\hbar^2/2I$ will be employed as the energy unit. We intend to analyze the quantum dynamics of the system in conditions of significant confinement of the rotors, and to this purpose we have selected the potential barrier as $u=300 (\hbar^2/2I)$.

The  contribution $ V^{(r)}$  of the system Hamiltonian has the purpose of producing a dynamical coupling between rotors by means of random interactions typical of material systems. Moreover, it assures the rational independence of the Hamiltonian $\hat H$ eigenvalues, property which does not hold in the presence of identical single particle Hamiltonians only. The random potential has been parameterized  as single particle contributions and interaction terms between pairs of rotors:
\begin{equation}
V^{(r)}(q_1, \dots, q_n)=\sum_{i=1}^nV^{(r)}_i(q_i)+\frac{1}{2}\sum_{i,j=1}^n(1-\delta _{i,j})V_{i,j}^{(r)}(q_i-q_j).
\label{rpot}
\end{equation}
Let us denote with $V(\theta )$ the periodic function representative of a single particle contribution, that is $V^{(r)}_i(q_i)$ for $q_i=\theta$, or of a two-particle interaction, that is $V_{i,j}^{(r)}(q_i-q_j)$ for $q_i-q_j=\theta$. By means of a gaussian random variable with null average and a  given variance $\sigma _V$, a random profile is easily generated for its discretized values $V_k:=V(\theta _k)$ at $(2L+1)$ equally spaced angles $\theta _k=2\pi k/(2L+1)$ for $k=0,1,\dots ,2L$. Standard algorithms can be employed to produce these random values of the function with statistical properties
\begin{equation}
\overline {V_k}=0, \text{ }\text{ }\text{ }\text{ }\text{ } \overline {V_k^2}=\sigma _V^2,
\end{equation}
where the average is referred to different realizations of the same coefficient. In order to recover a continuous function $V(\theta )$ from these random coefficients, we resort to a truncated Fourier decomposition 
\begin{equation}
V(\theta )=\sum_{l=-L}^L\tilde V_l e^{\imath l\theta},
\end{equation}
with its $(2L+1)$ components evaluated at the discretized angles
\begin{equation}
\tilde V_l=\frac  {1} {2L+1} \sum_{k=0}^{2L} V_k e^{-\imath l\theta _k}.
\end{equation}
Since an additive constant in the potential does not modify the quantum dynamical properties, a null value is attributed to the Fourier component $\tilde V_0$, this being equivalent to the constraint of a null angular average for functions $V(\theta )$.

\begin{figure}
\includegraphics[width=1\columnwidth, height=0.25\textheight]{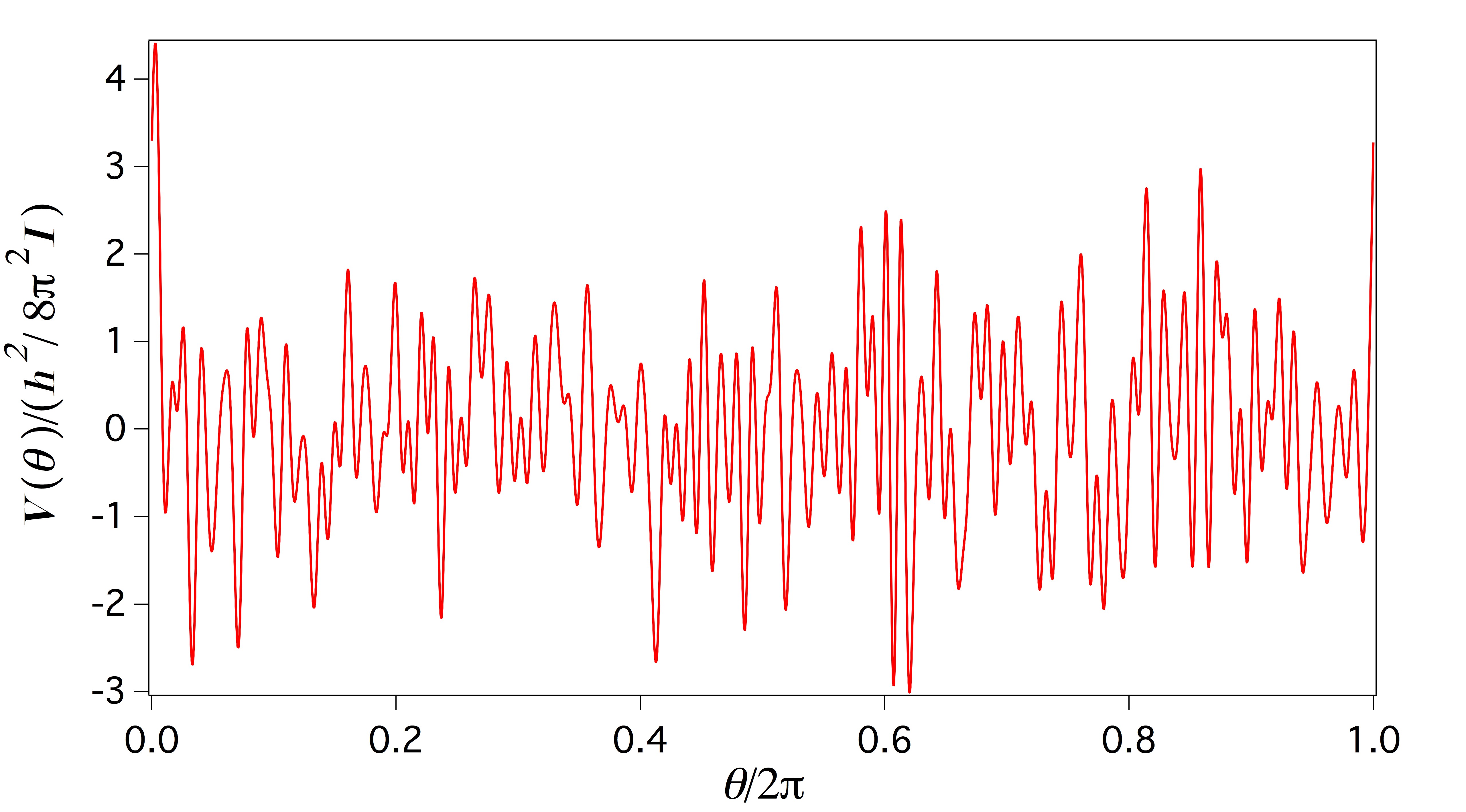}
\caption{\label{pot}Random potential characterized by the parameters $\sigma _V = \hbar ^2/2I$ and $L=100$.}
\end{figure}

In conclusion, the previous procedure allows the generation of these random angular functions for each contribution of Eq.~\eqref{rpot} on the basis of two parameters: the variance $\sigma _V$ and the number $(2L+1)$ of discretized angles. The variance $\sigma _V$ controls the strength of the random contribution $ V^{(r)} $ with respect to single particle Hamiltonians in Eq.~\eqref{hamiltonian}. In the following calculations we shall use an unitary value of this variance in the adopted energy units,  that is $\sigma _V = \hbar ^2/2I$. This corresponds to random potentials with a strength much smaller than the confining potential with a barrier height $u=300(\hbar ^2/2I)$. In this way, the random potential contribution has nearly a perturbation effect so that eigenfunctions and eigenvalues of the full Hamiltonian Eq.~\eqref{hamiltonian} preserve the main features deriving from single rotor contributions. The other parameter $L$ controls the size of the angular correlations in the potential, since it determines the distance between two adjacent discretized angles with uncorrelated values of the potential. In the calculations we shall use the value $L=100$ because it produces an highly random potential. An angular dependence resembling that of a noisy signal is evident form Fig.~\ref{pot} which displays the potential deriving from a particular realization of the $(2L+1)$ coefficents $V_k$ for $L=100$. For each of the contributions of Eq.~\eqref{rpot} an independent realization of the random potential $V(\theta )$ is employed.

\subsection{Numerical methods}

Before to discuss the results for the quantum dynamics of the model system, we summarize  in this section the numerical procedures employed in the calculations. They concern four main issues: i) the solutions for the single rotor Hamiltonian, ii) the eigenvalues and the stationary states for the system of interacting rotors, iii) the initial conditions and the time dependent wave-function and related properties, iv) the Bohm's trajectory.

\begin{figure}
\includegraphics[width=1\columnwidth, height=0.25\textheight]{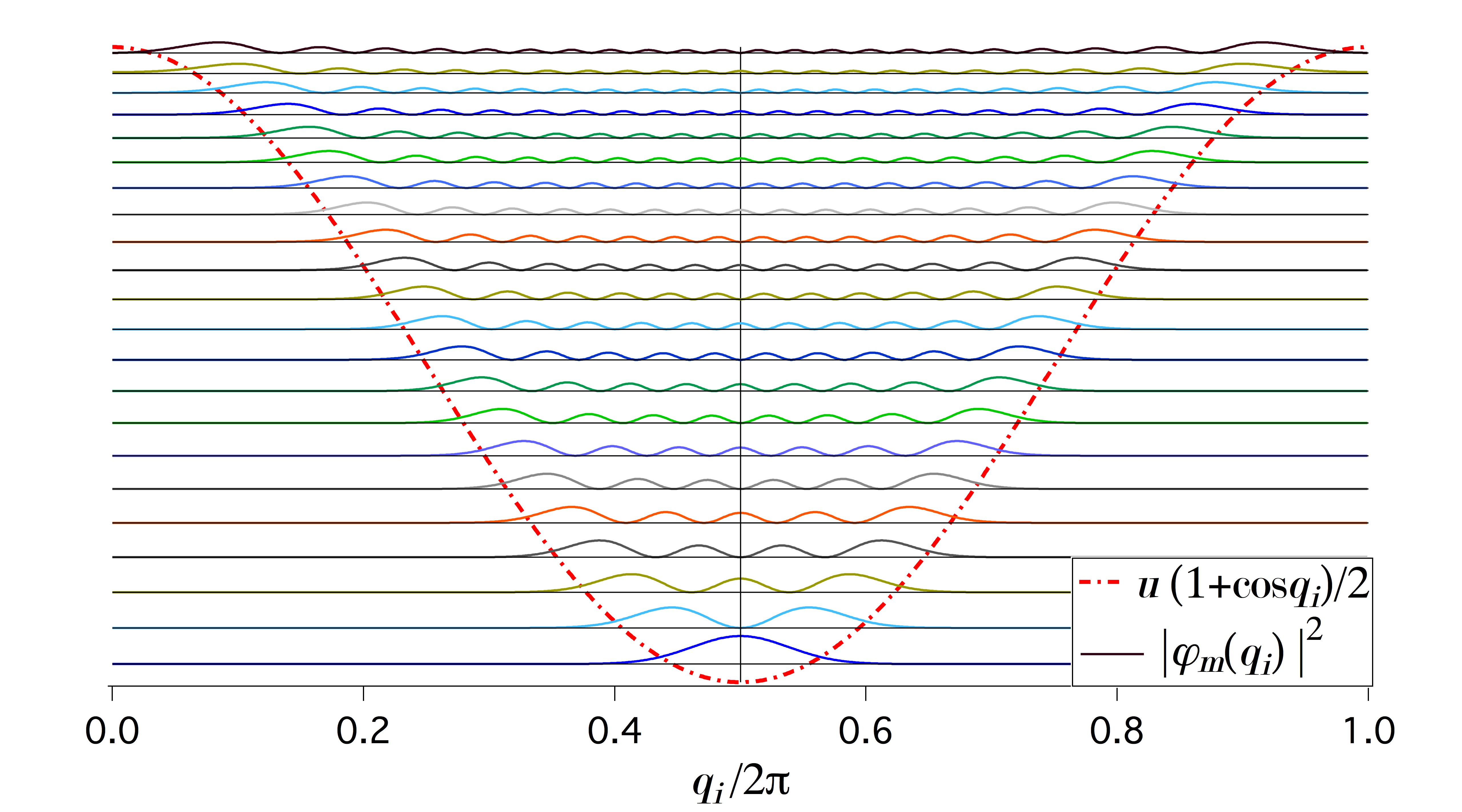}
\caption{\label{state}Squared modulus of eigenfunctions $\varphi_m(q_i)$ of the $H^{(0)}_i$ Hamiltonian with $u=300 (\hbar^2/2I)$ as the  intensity of the confining  potential $u(1+\cos(q_i))/2$}
\end{figure}

The eigenfunctions of the single rotor Hamiltonian Eq.~\eqref{Hrotor} are required because their tensorial products represent the most convenient basis for the numerical solution of the time independent Schroedinger Eq.~\eqref{tiSE} as long as the random potential $V^{(r)}$ is weak. Let us denote the eigenvalue problem for the single rotor as
\begin{equation}
\hat H_i^{(0)} \varphi _m(q_i)=\epsilon _m \varphi_m(q_i) ,
\label{eigenrotor}
\end{equation}
with eigenvalues ordered from below, $\epsilon _m \le \epsilon _{m+1}$, for $m=0,1,2,\dots$. Of course all the rotors have the same eigenvalues and eigenfunctions with the same functional form since they share the same Hamiltonian Eq.~\eqref{Hrotor}. In order to obtain the relevant eigenfunctions and eigenvalues with the required precision, we have generated the matrix representation of Eq.~\eqref{Hrotor} on the basis Eq.~\eqref{basis}
 for $|j|\le 20$, and diagonalized it by employing the software routine Armadillo, a \verb!C++! linear algebra library~\cite{armadillo}.  In Table~\ref{aval} we have reported the lower energy eigenvalues, and in Figure~\ref{state} the  profiles of the corresponding squared eigenfunctions $|\varphi _m(q_i)|^2$  with the eigenvalues as offset together with the confining potential.
In Table~\ref{aval} we have also included the harmonic oscillator eigenvalues resulting from the parabolic approximation $u(1+\cos q_i)/2\simeq u(q_i-\pi )^2/4$ of the rotor potential, in order to attest the differences with respect to purely harmonic quantum dynamics. Indeed for increasing 
levels the difference between the two eigen-energies clearly emerges. It should be mentioned that the numerical diagonalization of the single rotor Hamiltonian supplies not only the eigenvalues $\epsilon _m$, but also the eigenvectors, that is the coefficients for the expansion of the eigenfunctions $\varphi_m(q_i)$ on the basis of Eq.~\eqref{basis}. When, in the following, operations on single rotor eigenfunctions  $\varphi_m(q_i)$ are invoked, implicitly we refer to operations on these linear combinations which can be easily encoded in computer programs.

\begin{table}
\caption{\label{aval} Low energy eigenvalues $\epsilon _m $ of single rotor Hamiltonian for the potential barrier $ u=300 (\hbar^2/2I)$. The corresponding harmonic oscillator eigenvalues are reported between parenthesis. }
\begin{ruledtabular}
\begin{tabular}{cc}
$m$ & $\epsilon _m /(\hbar^2/2I)$  \\
$0$ & $8.597\ (8.660)$   \\
$1$ & $25.664\ (25.981)$  \\
$2$ & $42.472\ (43.301)$  \\
$3$ & $59.015\ (60.622) $  \\
$4$ & $75.286 \ (77.942)$  \\
$5$ & $91.278\ (95.263)$  \\
$6$ & $106.982\ (112.583)$  \\
$7$ & $122.390\ (129.904)$ \\
$8$ & $137.491\ (147.224)$ \\
$9$ & $152.275\ (164.545)$ \\
\end{tabular}
\end{ruledtabular}
\end{table}

Given the numerical solutions of Eq.~\eqref{eigenrotor}, one can employ the following basis  for the Hilbert space $\mathcal{H}$ of the overall system
\begin{equation}
\ket{l}=\bigotimes _{i=1}^n\ket{\varphi _{l_i}(q_i)},
\label{lbasis}
\end{equation}
where $l:=(l_1,l_2,\cdots,l_n)$, and each index $l_i$ identifies the eigenfunction $\varphi _{m}(q_i)$ of the corresponding $i$-th rotor with $m=l_i$. The basis elements $\ket{l}$ are eigenfunctions of the model system Hamiltonian in the absence of the random potential
\begin{equation}
\hat H^{(0)} \ket{l}=E^{(0)} _l \ket{l}, \text{ }\text{ }\text{ }\text{ }\text{ }E^{(0)} _l =\sum _{i=1}^n\epsilon _{l_i},
\end{equation}
and they are conveniently ordered according to the corresponding energies $ E^{(0)} _l$. As long as the random potential $V^{(r)}$ acts like a perturbation, the diagonalization of the full Hamiltonian is influenced mainly by the coupling between basis elements with nearby values of $  E^{(0)} _l$, and this allows an efficient truncation of the Hamiltonian matrix representation. In practice, one considers all the basis elements with $ E^{(0)} _l$ less than a given truncation energy cutoff  $ E^{(0)} _{tr}$. Then the matrix representation of the full Hamiltonian is generated in order to perform the diagonalization by means of the software Armadillo.

 \begin{figure}
\includegraphics[width=1\columnwidth, height=0.25\textheight]{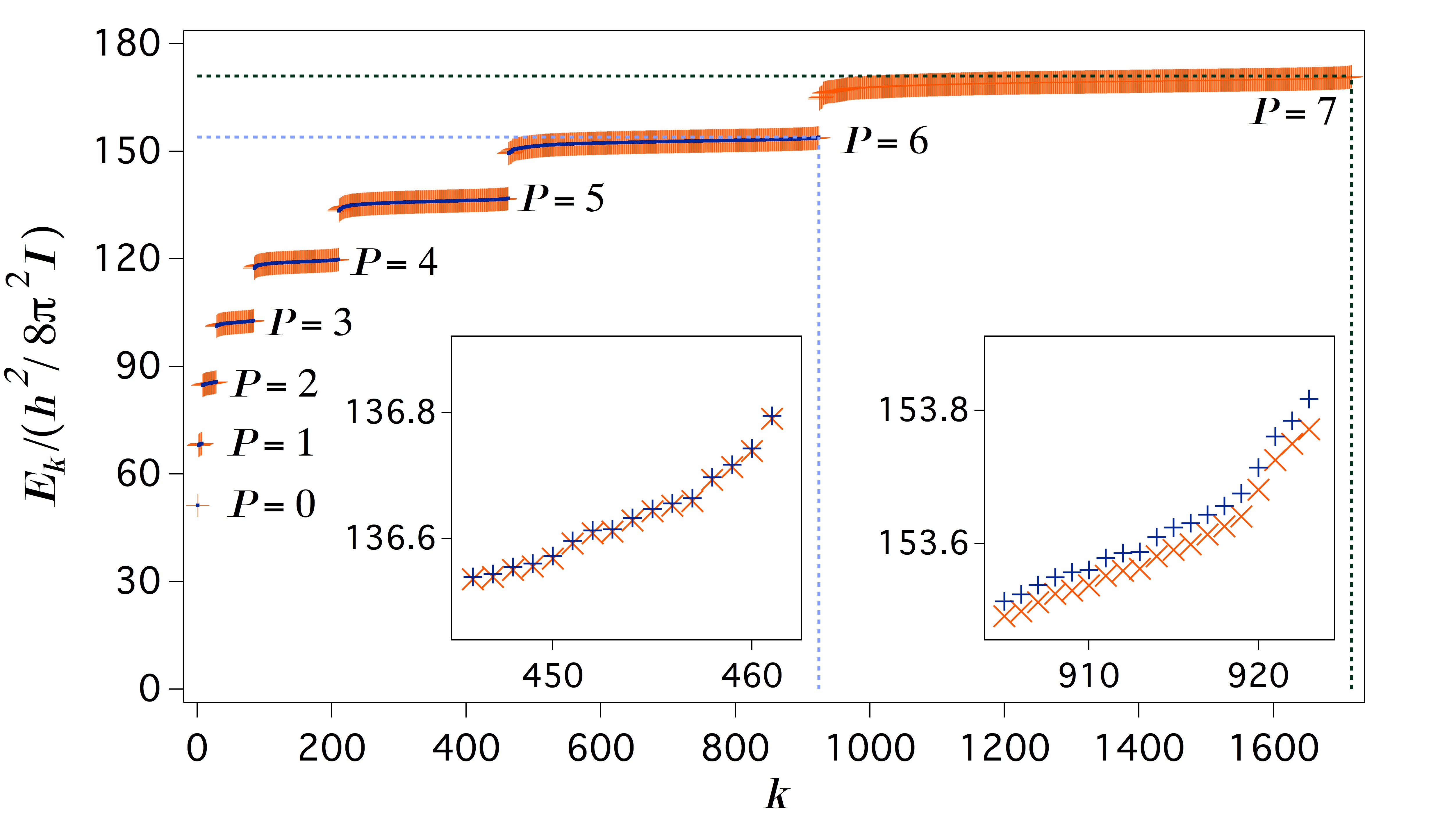}
\caption{\label{dos}Numerical results of energy eigenstates $E_k$ for two truncation parameters $ E^{(0)} _{tr}=154/(\hbar^2/2I)$ (blue crosses) and $ E^{(0)} _{tr}=171/(\hbar^2/2I)$ (red crosses). The states with comparable energy are labeled according to the polyad quantum number, from P=0 (ground state) to P=7. The last part of polyads $P=5$ and $P=6$ are magnified in the insets in order to show the effects of the truncation parameter. }
\end{figure}
The organization of these energy levels in well separated multiplets is evident in analogy to the polyads describing molecular vibrations (see~\cite{Krasnoshechekov2013, Herman2013} and references therein). Since in the harmonic approximation the oscillators for the confined rotors are degenerate, the polyad quantum number classifying the basis elements Eq.~\eqref{lbasis} is given as $P=\sum  _{i=1}^n {l_i}$  with values $P=0$ (ground state), $P=1$ (6 states), $P=2$ (21 states), and so on. As long the random potential $V^{(r)}$ is weak, the corresponding Hamiltonian eigenfunctions $\ket{E_k}$ are substantially reproduced by linear combinations of basis elements with a given polyad quantum number $P$, with only perturbational contributions from the other polyads. Therefore the polyad quantum number can be used to classify also the eigenvalue multiplets, as done in Fig.~\ref{dos}.

The comparison in Fig.~\ref{dos} of the numerical eigenvalues obtained with two values of parameter  $ E^{(0)} _{tr}$ allows one to evaluate the effects of matrix truncation. Notice that the chosen values of $ E^{(0)} _{tr}$ leads to a complete inclusion of the selected polyads in the truncated matrix representation. The results with matrix representation for 
$ E^{(0)} _{tr}=154/(\hbar^2/2I)$ ($N=924$), polyads from $P=0$ to $P=6$) will be employed as the reference for the calculation of time dependent properties of the model system. Their accuracy has been checked by comparison with the larger matrix obtained for  $ E^{(0)} _{tr}=171/(\hbar^2/2I)$ which includes a further polyad. Such a matrix enlargement, besides introducing new eigenvalues (i.e., the $P=7$ polyad), produces a change of about $0,04\%$ for the upper energy eigenvalues, and smaller variations for decreasing energy (see the insets of Fig.~\ref{dos}). Such a behavior agrees with the perturbational contribution by the random potential  $ V^{(r)}$: surely it has strong effects within a polyad in the presence of degenerate or nearly degenerate zero-order energies $ E^{(0)} _l$, but it has weak effects on the coupling of states belonging to different polyads with well separated values of   $ E^{(0)} _l$. These informations allows us to conclude that with the truncation parameter  $ E^{(0)} _{tr}=154 (\hbar^2/2I)$ ($N=924$) we get numerical results with errors at most of $0,04\%$. As a matter of fact the accuracy of the data employed in the calculation of time dependent properties is much better, as long as we shall use an active space including  up to $P=5$ polyad whose eigenvalues deviate from those obtained with the larger matrix by  $0,004\%$ at most. The final results of this computational task is the ensemble of eigenvalues $E_k$ and eigenfunctions $\ket{E_k}$, the latter specified as linear combinations of basis elements  Eq.~\eqref{lbasis} through coefficients $\braket{l|E_k)}$, for the time independent Schroedinger Eq.~\eqref{tiSE}. 

Once the eigenstates and the energy eigenvalues are obtained and the active space is identified on the basis of the cutoff energy $E_{max}$, the time dependent wave-function has to be evaluated. Thus the initial quantum state has to be chosen according to the set of  populations $P$ and the set of initial phases $\alpha (0)$ within the active space. Since the phases are homogeneously distributed~\cite{Emergence1}, they are simply selected at random within their domain. Also for the populations a random choice is performed but, in order to preserve their normalization, by means of suitable set of auxiliary parameters homogeneously distributed in the $(0,1]$ domain according to procedure discussed in Ref.~\cite{numRPSE1, numRPSE2, Beyond}. Given these initial conditions, the wave-function at an arbitrary time is specified as 
\begin{equation}
\ket{\Psi(t)}=\sum _{k=1}^N \sqrt{P_k}e^{-\imath[\alpha_k(0)+E_kt/\hbar]}\ket{E_k},
\label{wf}
\end{equation}
where $N$ is the dimension of the active space.
For the calculation of the reduced density matrix, reference is made to the first rotor, $q_S=q_1$, so that its matrix elements on the basis of single rotor eigenfunctions Eq. 
\eqref{eigenrotor} can be specified as 
\begin{equation}
\begin{split}
\sigma & _{m,m'}(t):= \braket{\varphi _m|\hat \sigma (t)|\varphi _{m'}}\\=&  \sum _{l,l'} \left( \prod _{i=2}^n \delta _{l_i,l_i'} \right)\delta _{l_1,m}\delta _{l'_1,m'} \braket {l|\Psi(t)} \braket {\Psi(t)|l'} =\\
=& \sum _{k,k'}  \sum _{l,l'} \left( \prod _{i=2}^n \delta _{l_i,l'_i} \right) \delta _{l_1,m}\delta _{l'_1,m'}\braket{l|E_k} \braket{E_{k'}|l{'}} \ \times \\
&  \times \sqrt{P_k P_{k'}} 
e^{-\imath[\alpha_{k}(0)-\alpha_{k'}(0) +(E_k-E_{k'})t/\hbar]}.
\end{split}
\end{equation}
The same equation with the constraint $k=k'$ in the summations on the r.h.s. can be employed to evaluate the elements  $\overline {\sigma}_{m,m'}$ of the equilibrium density matrix. Given the reduced density matrix, also the marginal quantum distribution of the subsystem (the first rotor) is recovered according to Eq.~\eqref{expdp}
\begin{equation}
p^S(q_S,t)=\sum _{m,m'}\sigma  _{m,m'}(t) \varphi _m(q_S) \varphi _{m'}^{*}(q_S),
\label{QuantDis}
\end{equation}
and the equilibrium distribution as well by inserting the equilibrium density matrix elements
\begin{equation}
p^{S,eq}(q_S)=\sum _{m,m'} \overline {\sigma}_{m,m'}  \varphi _m(q_S) \varphi _{m'}^{*}(q_S).
\label{EqQuantDis}
\end{equation}
By specifying the eigenfunctions $\ket{E_k}$ in Eq.~\eqref{wf} as linear combinations of the basis functions Eq.~\eqref{lbasis}, one gets for a given time the explicit dependence on the coordinates of the wave function, $\Psi (q,t)$, and of both the amplitude $R(q,t)$ and the phase $S(q,t)$ as well.

For the  computation of the trajectory of the rotors, we adopted the Runge-Kutta method~\cite{numericalrecipes} at the $4^{\text{th}}$ order to solve the Bohm equation of motion Eq.~\eqref{bohmeq}. We employed  a time step $\Delta t= 0.01(4\pi I/\hbar)$ that assures a good approximation to the calculated trajectory from the point of view of its statistical properties. In particular we have evaluated the correlation function $G(t)$ of the rotor angle $Q_S$
\begin{equation}
G(\tau ):=\overline {\Delta Q_S(t)\Delta Q_S(t+\tau )},
\label{CF}
\end{equation}
with $\Delta Q_S(t)=Q_S(t)-\overline{Q_S(t)}$, that we calculate from the discretized time average along the trajectory:
\begin{equation}
G(\tau )\simeq \frac{1}{M} \sum_{j=0}^{M}  \Delta Q_S(j\Delta t)\Delta Q_S(j\Delta t+\tau ),
\label{CFnum}
\end{equation}
where $M$ is the number of sampling points that depends on the length of the examined trajectory.

Finally  the distribution $w^{S,eq}(q_S)$ of the rotor coordinate along its trajectory has to be evaluated. In practice we have calculated its discretized counterpart by dividing the domain $0\le Q_S<2\pi$ of the rotor angle into $10^4$ equally spaced intervals. The probability density is recovered from the fraction of time spent by the rotor in each interval during its evolution. In order to check that the length of the trajectory is sufficient, we have verified that the resulting distribution is not modified by a further evolution. 

One might wonder whether the numerical procedure for the calculation of the Hamiltonian eigenvalues and eigenfunctions, which provides always approximate results, affects the behavior of the computed trajectory. If this is the case, then in the comparison between the quantum distribution $p^{S,eq}(q_S)$ for the subsystem and the distribution $w^{S,eq} (q_S) $ on the Bohm's coordinate, one should consider explicitly the influence of the errors introduced by the numerical diagonalization. Let us denote with $E_k^{(app)}$ and $\ket {E_k^{(app)}}$ the approximate eigenvalues and eigenfunctions computed numerically. The computed wave-function derives from the linear combinations of these approximate eigenfunctions, and it is a solution of the Schroedinger equation for the Hamiltonian 
\begin{equation}
\hat H^{(app)}:=\sum _{k=1}^N \ket {E_k^{(app)}} E_k^{(app)}  \bra {E_k^{(app)}}
\end{equation}
instead of the assumed model Hamiltonian Eq.~\eqref{hamiltonian}. Correspondingly the phase function $S(q,t)$ and the resulting Bohm's trajectory is exact for the quantum problem described by the Hamiltonian $\hat H^{(app)}$. In conclusion, the unavoidable errors introduced by the numerical diagonalization are formally equivalent to a slight modification of the system Hamiltonian. Of course the Bohm's trajectory is also affected by the numerical errors in the integration of the differential equation~\eqref{bohmeq}, but their effects can be easily controlled by checking that the coordinate distribution $w^{S,eq} (q_S) $, and the coordinate correlation function Eq.~\eqref{CF} as well, does not change by decreasing the integration time step.

\subsection{Dynamical properties}
The selected model system and its Hamiltonian with $ u=300 (\hbar^2/2I)$ as the barrier of the confining potential, is compatible with different thermal states depending on the cut-off energy $E_{max}$ of the RPSE. We have selected the value $E_{max}=139(\hbar^2/2I)$ which corresponds to an active space of dimension $N=462$ including the polyads from $P=0$ to $P=5$ and exluding the other polyads (see Fig.~\ref{dos}). With such a choice we deal with a state having a significant distribution between the ground state and the excited states of the single rotor, as witnessed by the subsystem reduced density matrix represented on the basis of single rotor eigenfunctions $\varphi_m(q_S)$ of Eq.~\eqref{eigenrotor}. The equilibrium reduced density matrix calculated according to the methods illustrated in the previous section is nearly diagonal, and the diagonal components are reported in Table~\ref{drdm}. The calculated off-diagonal elements $\overline {\sigma}_{m,m'}$ are less than 1/1000 in magnitude with respect the associated diagonal elements $\overline {\sigma}_{m,m}$ and $\overline {\sigma}_{m',m'}$. The decrease of the diagonal elements $\overline {\sigma}_{m,m}$ with the single rotor energy $\epsilon _m$ (see Table~\ref{drdm}) might suggest a canonical form $\overline {\sigma}_{m,m} \propto \exp (-\beta \epsilon _m) $ but this is not the case. In order to provide evidences about it, we have derived the hypothetical canonical  thermal coefficient $\beta =0.0376 / (\hbar^2/2I)$ from the ratio  $\overline {\sigma}_{1,1} /  \overline {\sigma}_{0,0}$ and then the corresponding elements of the canonical density matrix, which are reported between parenthesis in Table~\ref{drdm}. The deviations with respect to the numerical values of  $\overline {\sigma}_{m,m}$ clearly emerge, particularly for the upper energy elements, and this points out that the size of our model system (six interacting rotors) is not large enough to ensure the thermodynamic limit. On the other hand the rather small differences for the more populated states (say, for $m=0,1,2$) allows us to drawn the conclusion that the state of the system resembles that of thermodynamic equilibrium.  Furthermore, the significant mixing of single rotor eigenstates clearly emerges from the data of Table~\ref{drdm}, since excited states (i.e., the elements with  $m\not= 0$) contribute by nearly 50\% to the reduced density matrix.

\begin{table}
\caption{\label{drdm} Diagonal elements of the equilibrium reduced density matrix, with their canonical values reported between parenthesis.}
\begin{ruledtabular}=
\begin{tabular}{cc}
$m$ & $\overline {\sigma}_{m,m}$ \\
$0$ & $0.536 \ (0.475)$   \\
$1$ & $0.282\ (0.250)$  \\
$2$ & $0.127\ (0.133)$  \\
$3$ & $0.0431\ (0.0712) $  \\
$4$ & $0.0122 \ (0.0386)$  \\
$5$ & $5.15 \ 10^{-4}\ (0.0211)$  \\
$6$ & $3.61\  10^{-7}\ (0.0117)$  \\
\end{tabular}
\end{ruledtabular}
\end{table}

\begin{figure}
\includegraphics[width=1\columnwidth, height=0.25\textheight]{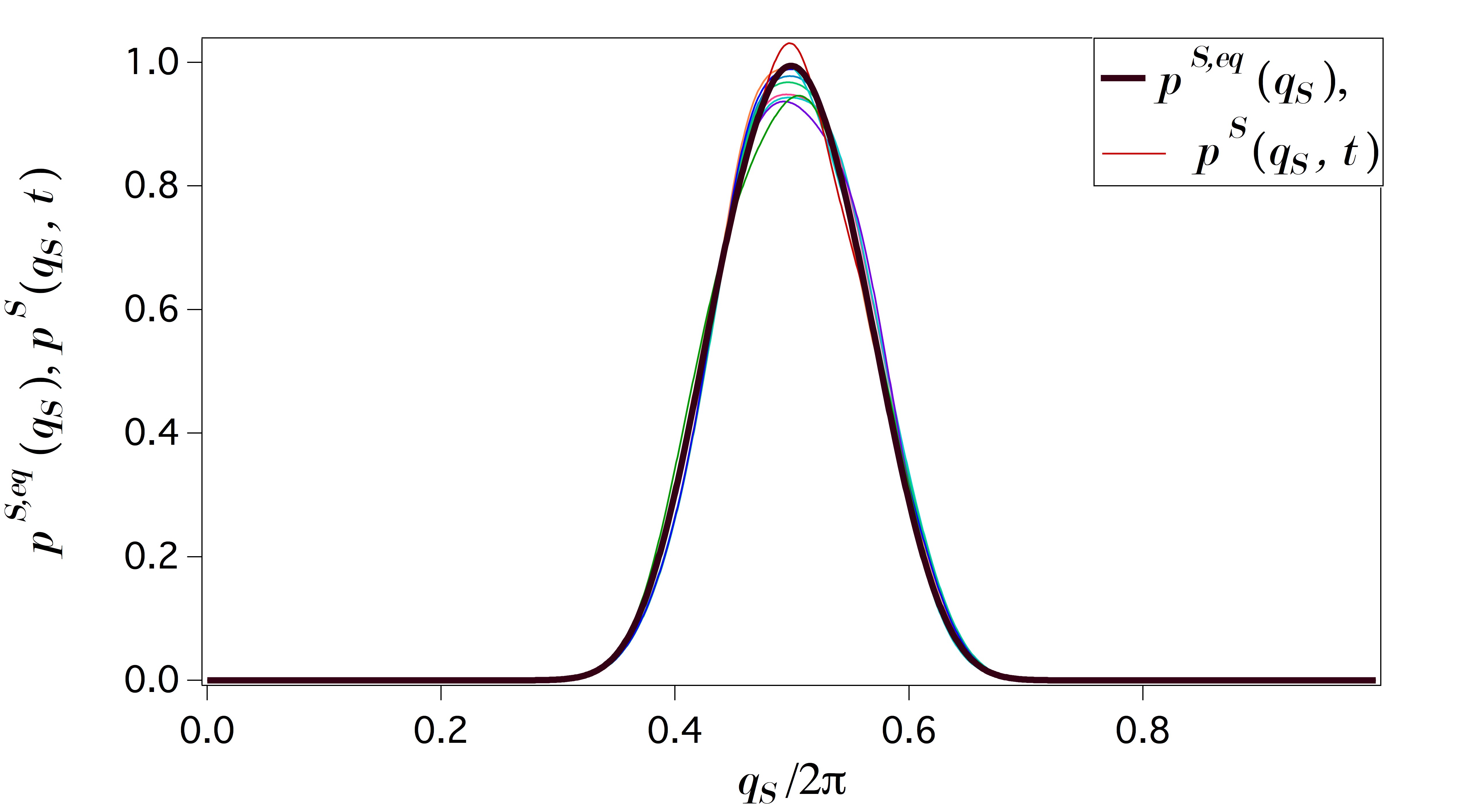}
\caption{\label{dqt}Equilibrium marginal density distribution  $p^{S,eq}(q_S)$ (black thick line) and marginal density distributions  $p^{S}(q_S,t)$ (colored thin lines) at some selected times. The marginal distributions are referred to the first of the 6 rotors in our model system.}
\end{figure}

As explained in the previous section, the instantaneous reduced density matrix allows the calculation of the time dependent quantum distribution $p^S(q_S,t)$ of the subsystem (the first rotor). The profiles of such a distribution are reported in Fig.~\ref{dqt} for a selected sample of times.
As the reference for the visualization of its change with the time, in the same Figure we have plotted also the equilibrium quantum distribution $p^{S,eq}(q_S)$ calculated according to equilibrium reduced density matrix  $\overline {\sigma}_{m,m'}$. In Section II we have shown that in the thermodynamic limit, i.e., when the number of interacting components is large enough, the fluctuations of $p^S(q_S,t)$ become negligible and then $p^{S,eq}(q_S)$ would reproduce the quantum distribution function at all times. The data in Fig.~\ref{dqt} clearly show that this is not the case in our model system as long as time dependent deviations from $p^{S,eq}(q_S)$ are evident. On the other hand these deviations have a comparably low magnitude, so that the equilibrium distribution $p^{S,eq}(q_S)$ can be considered as representative, at least approximately, of the instantaneous quantum distribution $p^S(q_S,t)$. 

\begin{figure}
\includegraphics[width=1\columnwidth, height=0.25\textheight]{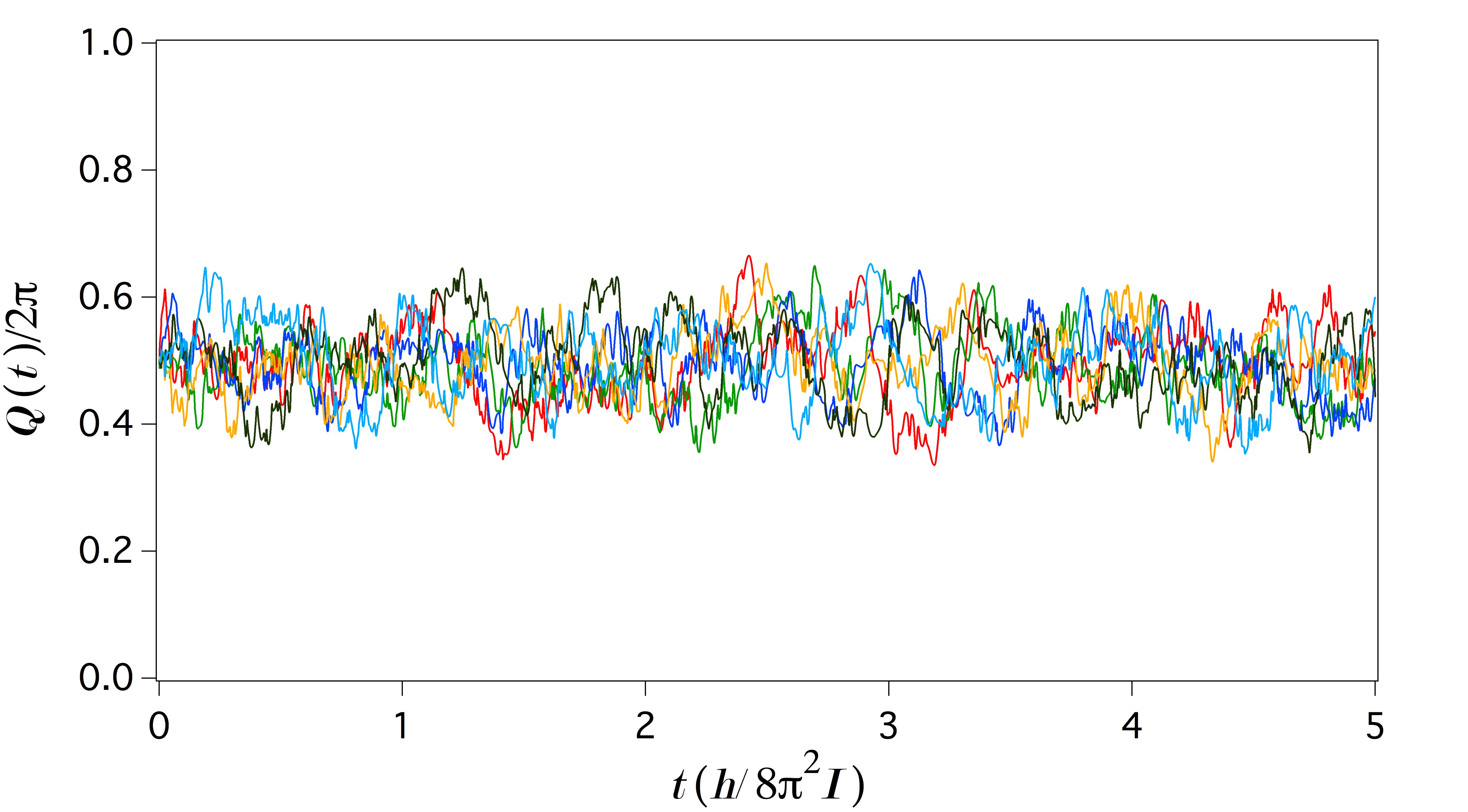}
\caption{\label{tra}Time evolution of the Bohm coordinates (drawn with different colors) of the six rotors of the model system.}
\end{figure}

\begin{figure}
\includegraphics[width=1\columnwidth, height=0.25\textheight]{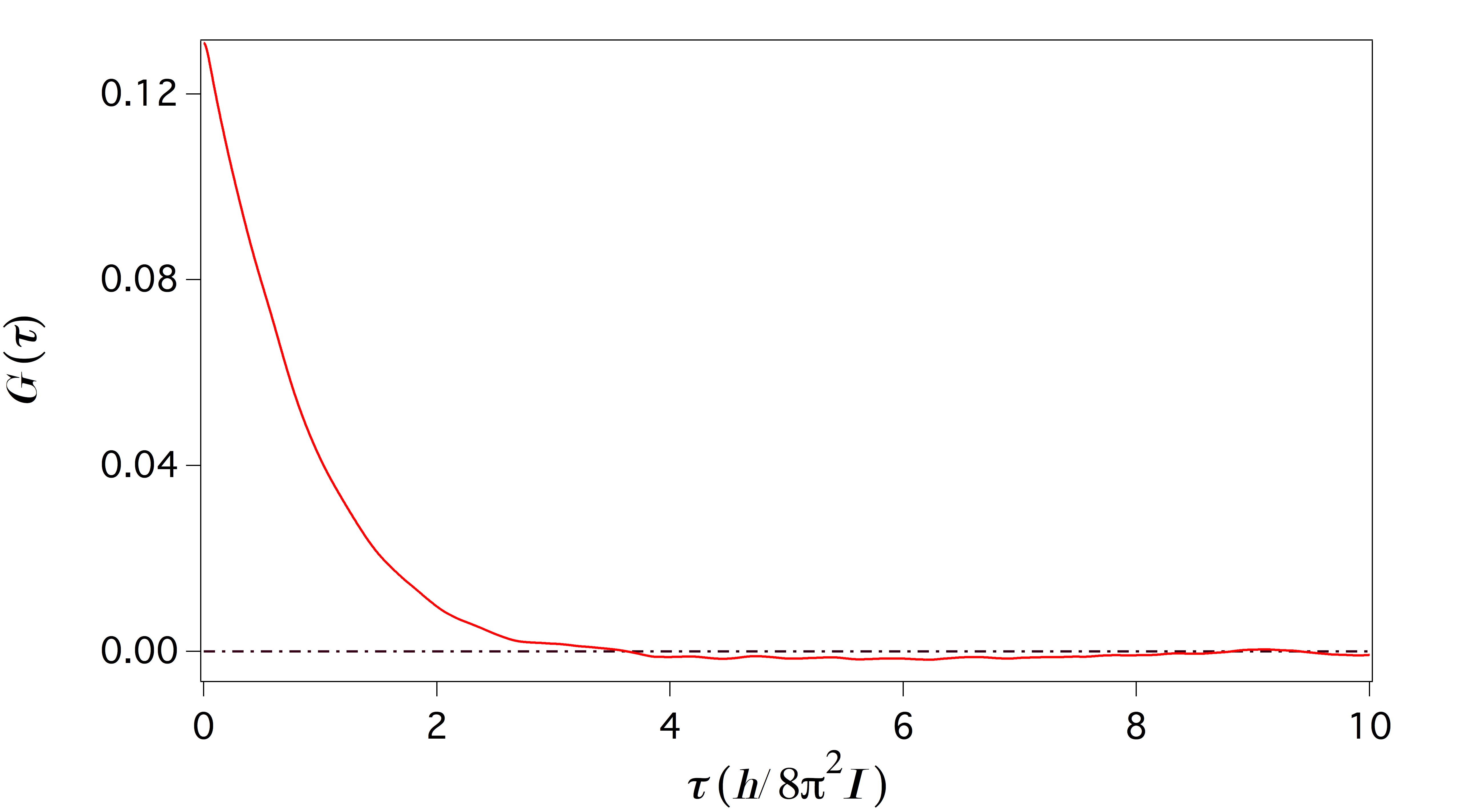}
\caption{\label{corr}Correlation function $G(\tau)$ of the first rotor coordinate.}
\end{figure}

Having characterized the main quantum properties of the subsystem, we examine now the behavior of the Bohm's coordinates. By employing the procedure illustrated in the previous section, the  trajectories of the angular coordinates $Q_i(t)$ of the six rotors have been computed according to Eq.~\eqref{bohmeq} by choosing $Q_i(0)=\pi$ as initial conditions in correspondence of the bottom of the rotor confining potential. In Fig.~\ref{tra} we have represented with different colors the trajectories of all the rotors within the time window $0\le t/ (4\pi I/\hbar) \le 5$.
 Each rotor coordinate follows a strongly confined dynamics with limited excursions about the potential minimum. The time evolution of each rotor coordinate seems that of a fluctuating signal loosing correlation with time, somehow like in the brownian motion.  To verify this feature, we have computed the correlation function $G(\tau )$  Eq.~\eqref{CF} which is displayed in Fig.~\ref{corr}.
As expected on the basis of the behavior of the trajectories, the correlation vanishes with a rather short correlation time of order $\tau _c /(4\pi I/\hbar) \simeq 4 $ supporting the analogy with brownian motion. It appears that the phase function $S(q,t)$ due to the wave-function generates a fluctuating evolution of the Bohm's coordinates, which leads to a fast loss of correlation. 

\begin{figure}
\includegraphics[width=1\columnwidth, height=0.25\textheight]{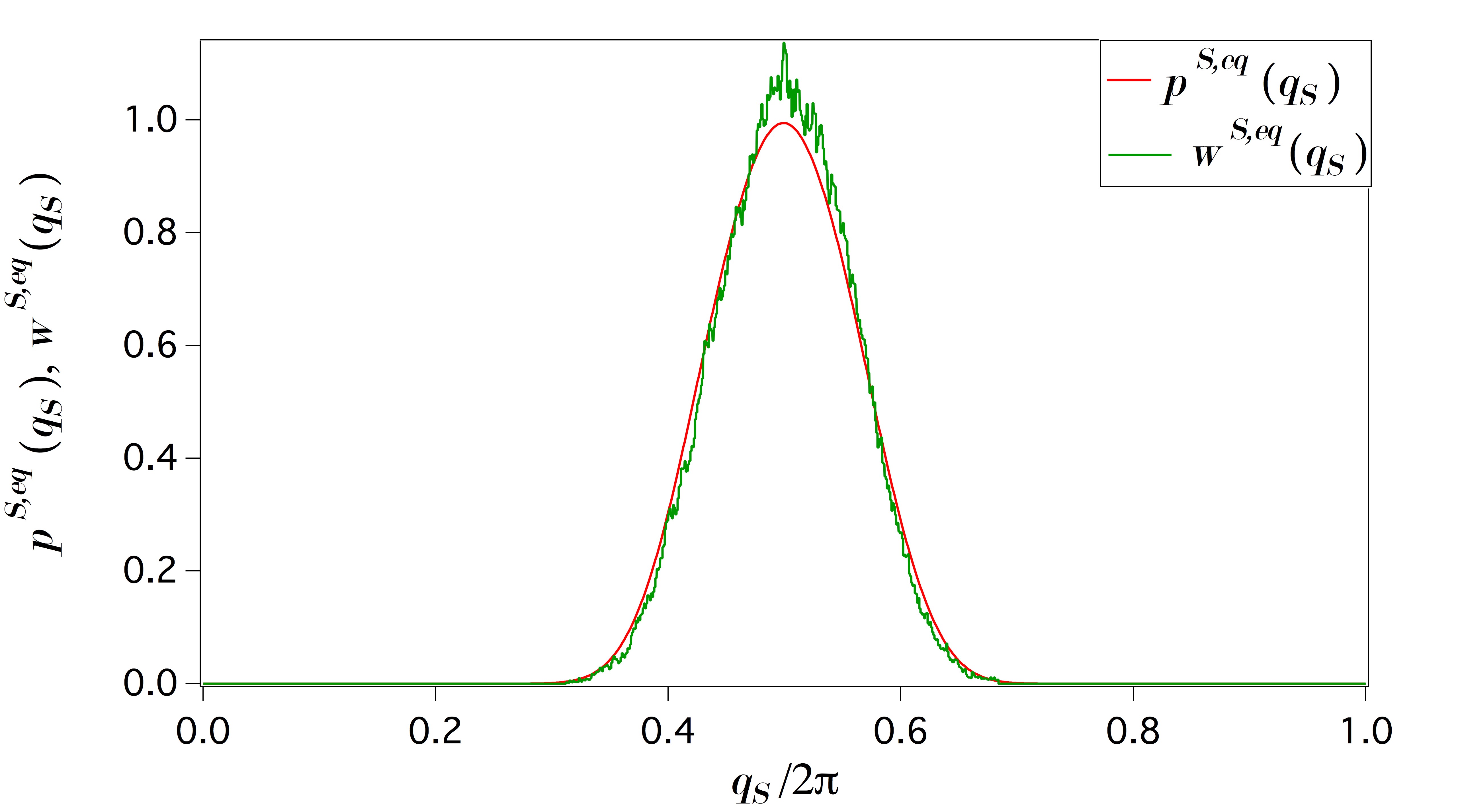}
\caption{\label{guess}Equilibrium marginal quantum distribution $p^{S,eq} (q_S)$ and marginal density distribution $w^{S,eq}(q_S)$ of the Bohm coordinate for the first rotor.}
\end{figure}

If the Bohm's coordinate $Q_S$ of the subsystem is considered as a stochastic process, than its properties are naturally characterized by the correlation function Eq.~\eqref{CF} and its equilibrium distribution $w^{S,eq} (q_S)$. The expected confinement of the rotor angle clearly emerges from such a distribution which is displayed in Fig.~\ref{guess}.
In the same Figure we have plotted also the equilibrium quantum distribution for the sake of comparison. The two equilibrium distributions, the one $p^{S,eq} (q_S)$ deriving from the evolution of the wave-function, and the other $w^{S,eq} (q_S)$ calculated from a single Bohm's trajectory, result to be very close
\begin{equation}
w^{S,eq} (q_S) \simeq p^{S,eq} (q_S).
\label{corrispondenza}
\end{equation}
Notice that the loss of correlation along the trajectory implies that the distribution $w^{S,eq} (q_S)$ is independent of the choice of the initial values $Q(0)$ of the Bohm's coordinates.

\begin{figure}
\includegraphics[width=1\columnwidth, height=0.25\textheight]{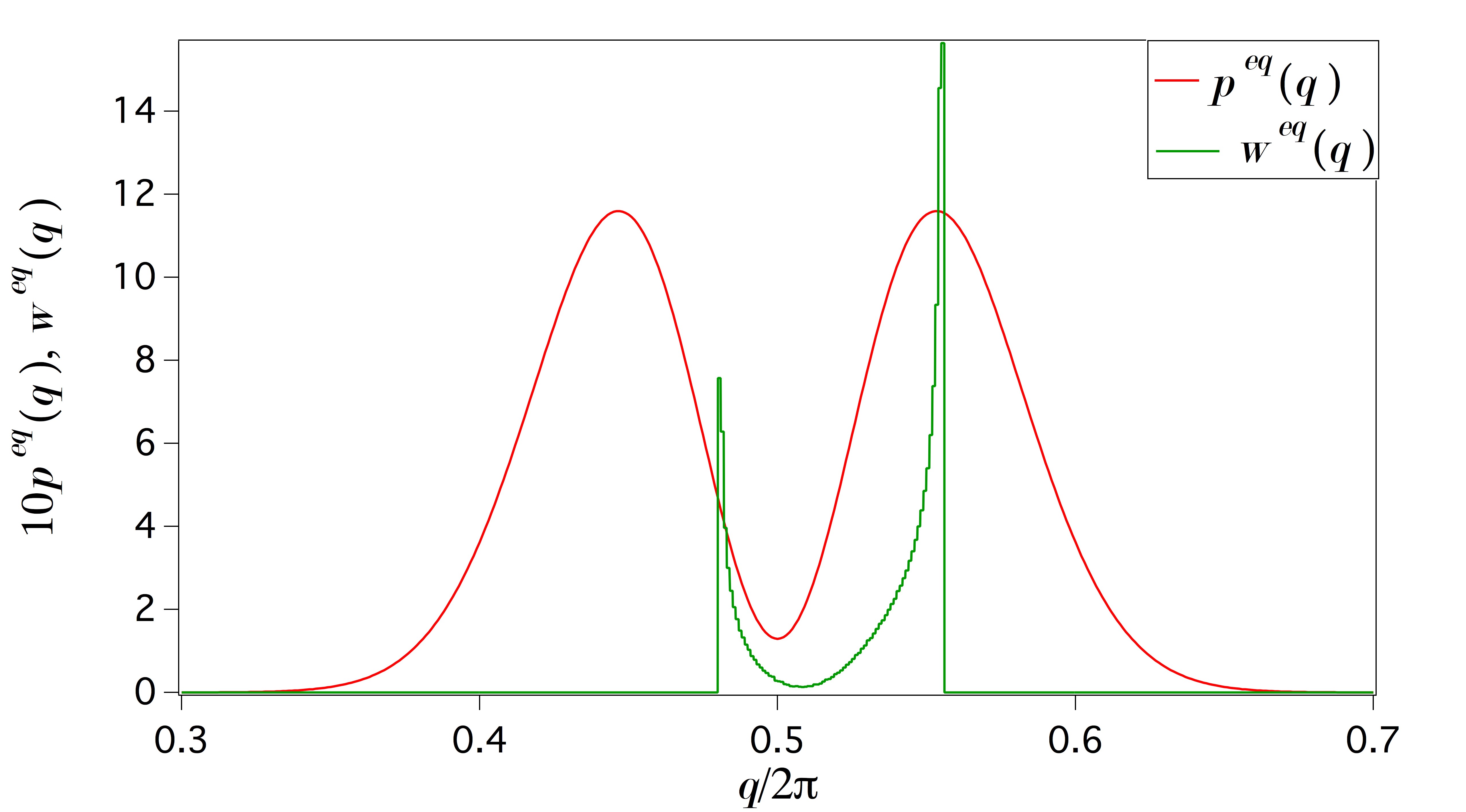}
\caption{\label{noguess}Quantum equilibrium distribution $p^{eq} (q)$ and the distribution $w^{eq}(q)$ of the Bohm coordinate for the  model system with a single  rotor.}
\end{figure}

It should be stressed that the  correspondence Eq.~\eqref{corrispondenza} cannot be considered as a general property for  all quantum systems. Indeed one can use a single rotor system as a counterexample where Eq.~\eqref{corrispondenza} does not hold. If the same previous procedure is applied to an isolated confined rotor, with the same potential of our model system, by choosing an active space of dimension $N=2$ in order do deal with a wave-function with a nearly 50\% probability of the ground state like for the reduced density matrix of Table~\ref{drdm}, one obtains two very different equilibrium distributions like those displayed in Fig.~\ref{noguess}.
It should be mentioned that the asymmetry on the profile of $w^{eq} (q)$  derives from the difference of the randomly chosen populations of the two quantum states. We conjecture that the correspondence Eq.~\eqref{corrispondenza}  found in our model system is a consequence of the multi-particle interactions which are absent in the single rotor system.

Finally we emphasize that the correspondence~\eqref{corrispondenza} is not an accidental result of particular conditions employed for the calculation in our model system. As a matter of fact we have similar evidences of the correspondence from calculations in other conditions, for instance by using a confining potential with lower strength or even in the absence of the confining potential.


\section{Bohm coordinates as Markov stochastic variables\label{just}}
As reported in the previous section, explicit calculations with a many body model system suggest that, even by considering pilot wave theory at the level of single Bohm's trajectory, a correspondence  exists according to Eq.~\eqref{corrispondenza}  between  probability density determined by the wave-function and the coordinate distribution derived from the trajectory. This is an important result since it allows a connection between the standard quantum theory and the geometrical description of system evolution through a trajectory, without the need of the Born's rule and the corresponding swarm of trajectories. On the other hand, such a connection has a methodological role different from the Born's rule. In particular we emphasize the following three features. 
1) It is a correspondence concerning the subsystem only, while the Born's rule Eq.~\eqref{Bornt} deals with the overall isolated system. 2) Its validity has to be restricted to the case of negligible fluctuations on the quantum distribution Eq.~\eqref{pista} which then can be replaced  by the equilibrium quantum distribution Eq.~\eqref{pmedia}. Only in this case the quantum distribution becomes time independent and, therefore, it can be compared with  Bhom's  coordinate distribution which, by definition, is time independent. 3) Such a  correspondence in general is not exact, since there are evident counterexamples (Figure~\ref{noguess}). The calculation results simply suggest that quantum distribution and Bohm's trajectory coordinate distribution are close in suitable conditions. On the contrary the Born's rule Eq.~\eqref{Bornt} is exactly verified once the third Bohm's assumption Eq.~\eqref{Born0} is introduced. 
 
Besides these considerations, an important issue naturally arises: can the correspondence Eq.~\eqref{corrispondenza} find a support beyond the evidences resulting from  calculations with specific model systems? In other words, can  Eq.~\eqref{corrispondenza} be derived under particular conditions?  A positive answer is found if the evolution of subsystem coordinate in the single Bohm's trajectory follows a stationary Markov process for a stochastic variable~\cite{gardiner}. A stationary Markov process is completely characterized by the equilibrium distribution $w^{S,eq} (q_S)$ and the conditional probability distribution $w^S (q_{S,0}|q_S,\tau )$. The former is obtained from the sampling of subsystem coordinate $Q_S(t)$ along a single Bohm's trajectory, as we have done in our model system. The latter requires the sampling of the correlation of coordinates $Q_S(t)$ and $Q_S(t+\tau )$ at two times separated by $\tau$, and it should satisfy the constraint of correlation loss at long enough times:
\begin{equation}
\lim_{\tau\rightarrow +\infty}w^S(q_{S,0}|q_{S}, \tau)=w^{S,eq}(q_S).
\label{lostcorrelation}
\end{equation} 

The distributions characterizing the Markov process observed in a single trajectory, can be used to describe also  the probability density arising from an ensemble of trajectories. Let us denote with $\rho ^S (q_S,t)$ the probability density on the coordinate for such an ensemble of trajectories. Given the initial distribution  $\rho ^S (q_S,0)$, the probability density at any time can be evaluated on the basis of the correlation function $w^S (q_{S,0}|q_S,\tau )$,            
\begin{equation}
\rho ^S (q_S,t)=\int  \mathrm{d}q_{S,0}\text{ } \rho ^S(q_{S,0},0) w^S (q_{S,0}|q_S,t)  ,
\label{rot}
\end{equation} 
and, according to Eq.~\eqref{lostcorrelation}, it will relax to the equilibrium distribution $w^{S,eq}(q_S)$ at long enough times
\begin{equation}
\lim_{\tau\rightarrow +\infty}\rho ^S (q_S,t)=w^{S,eq}(q_S).
\label{limro}
\end{equation}
Let us now recognize the conditions under which the distribution on the trajectories is stationary, that is $\rho ^S (q_S,t)$ is time independent. Stationarity means that the equivalence in Eq.~\eqref{limro} must be verified at all times,
\begin{equation}
\rho ^S (q_S,t)=w^{S,eq}(q_S),
\label{statro}
\end{equation}
and, therefore, the initial distribution $\rho ^S (q_S,0)=w^{S,eq}(q_S)$ is the unique condition leading in Eq.~\eqref{rot} to a time independent distribution.

Let us  apply these results to the ensemble of trajectories generated according to the third Bohm's assumption, and with the further conditions: a) stationary quantum distribution for the subsystem, $p^S (q_S,t)=p^{S,eq}(q_S)$, b) stationary Markov process for subsystem coordinate in a Bohm's trajectory. Then the subsystem probability density,  according to the Born's rule Eq.~\eqref{Bornt}, is equivalent to the stationary quantum distribution
\begin{equation}
\rho ^S (q_S,t)=p^S (q_S,t)=p^{S,eq}(q_S),
\label{statro2}
\end{equation}
 because of condition a).
On the other hand, because of condition b), the same probability density can be computed according to Eq.~\eqref{rot} with $\rho ^S (q_S,0)=p^{S,eq}(q_S)$
as the initial distribution. However, such a probability density has only one stationary form specified by Eq.~\eqref{statro}. One can conclude that, as long as the two conditions a) and b) are satisfied and, therefore, Eqs.~\eqref{statro} and~\eqref{statro2} are holding simultaneously,   for the subsystem the quantum equilibrium distribution and the coordinate distribution in a Bohm's trajectory are equivalent, 
\begin{equation}
p^{S,eq}(q_S)=w^{S,eq}(q_S).
\label{equivalence}
\end{equation}
This is the important result of the previous analysis which, however, is conditioned by the validity of assumptions a) and b).  In section II we have analyzed the fluctuations of the quantum distribution for the subsystem, by showing that they vanish in the limit of an infinite size environment. This points out that in finite but large enough systems, condition a) is satisfied only approximately, and the same type of validity should be attributed to the  equivalence Eq.~\eqref{equivalence}. At this stage a specific analysis about the general validity of the description of subsystem coordinate as a Markov process is lacking even if, in analogy to classical brownian motion, one might conjecture that such a feature is determined by the coupling amongst several degrees of freedom. On the other hand the model results reported in the previous section suggest that for systems characterized by random interactions amongst its components, the subsystem evolution leads to distributions approximating Eq.~\eqref{equivalence}.


\section{Conclusion\label{concl}}
We have considered the system of six confined rotors as a model to test the representation of quantum systems by the single Bohm's trajectory. For the subsystem identified with one rotor, the others playing the role of  the environment, we have found the following main results from the numerical solution of the pilot wave theory: 1) the marginal quantum distribution derived from the wave-function is nearly stationary, 2) the Bohm's coordinate evolves like a randomly fluctuating signal with a clear loss of correlation with the time, 3) the rather close correspondence between the marginal quantum distribution and the distribution of the Bohm's coordinate. We stress the interest of the last result in relation to the methodological status of the pilot wave theory. If a correspondence exists between the quantum distribution derived from the wave-function and the distribution of the Bohm's coordinate along a trajectory, albeit at the level of the subsystem, then the ensemble of trajectories together with the Born's rule for their initial distribution is not mandatory to establish a connection between standard quantum theory and the particle's configuration of the pilot-wave theory. A more direct picture of material systems would then be derived on the basis of a single realization of both the quantum state  (the wave-function) and the particle's configuration (the Bohm's trajectory).

On the other hand we emphasize that such a correspondence is presently a conjecture even if supported by the numerical results for a particular model system. The existence of conditions assuring the validity of the conjecture remains still an open issue. We were able to verify it only under the hypothesis that the Bohm's coordinate behaves as a Markov stochastic process, as shown in the previous section. 

Finally we would like to comment on the implications of the near stationarity of the marginal quantum distribution for the subsystem, as shown in Fig.~\ref{dqt}. This is strictly a consequence of the vanishing of fluctuations of the reduced density matrix in the thermodynamic limit as analyzed in Ref.~\cite{flureddensmatrix}. A direct relation exists also with the typicality analyzed in Ref.~\cite{bartsch2009} even if in those theories the effects of fluctuations in time are not separated from the distribution within the ensemble. At any rate, a static picture of the subsystem properties would be implied from these results, at odds with the opposite image of an ever fluctuating world as suggested by the classical mechanics. We think that the pilot wave theory, in the single trajectory approach, leads to a solution of these contradictory representations. Indeed, as in displayed in Fig.~\ref{tra}, the fluctuating evolution of the Bohm's coordinate, very much like for a confined Brownian particle, results compatible with the nearly stationarity of the marginal quantum distribution.

\begin{acknowledgements}
The authors acknowledge the support by Univesit\`a degli Studi di Padova through $60\%$ grants.
\end{acknowledgements}

\bibliography{Art1_mybib}

\end{document}